\documentclass[11pt,journal,twoside,onecolumn]{IEEEtran}

\usepackage{psfrag,epsfig,graphics}
\usepackage{amsmath,amssymb,multirow}
\usepackage{subfigure,url}
  \usepackage[below]{placeins}
  \usepackage{cite}
  \usepackage[ruled,vlined]{algorithm2e}
  
  \usepackage[utf8]{inputenc} 
\usepackage[T1]{fontenc}
  \usepackage{balance}

  \usepackage{relsize}

\usepackage{color}  
\definecolor{agreen}{rgb}{0,0.5,0}

\def\R{\ensuremath{{\mathrm{I\!R}}}}

\newcommand{\cb}[1]{{\boldsymbol{#1}}}
\newcommand{\cp}[1]{\ifmmode {\mathcal{#1}}\else ${\mathcal{#1}}$\fi}

\newcommand{\beps}{\boldsymbol{\epsilon}}
\newcommand{\bpsi}{\boldsymbol{\psi}}
\newcommand{\bphi}{\boldsymbol{\phi}}

\newcommand{\btheta}{\boldsymbol{\theta}}

\newcommand{\bu}{\boldsymbol{u}}
\newcommand{\bp}{\boldsymbol{p}}
\newcommand{\bs}{\boldsymbol{s}}

\newcommand{\bg}{\boldsymbol{g}}
\newcommand{\br}{\boldsymbol{r}}

\newcommand{\bw}{\boldsymbol{w}}
\newcommand{\bx}{\boldsymbol{x}}

\newcommand{\bv}{\boldsymbol{v}}

\newcommand{\bX}{\boldsymbol{X}}

\newcommand{\bB}{\boldsymbol{B}}
\newcommand{\bC}{\boldsymbol{C}}
\newcommand{\bD}{\boldsymbol{D}}
\newcommand{\bK}{\boldsymbol{K}}

\newcommand{\bG}{\boldsymbol{G}}
\newcommand{\bP}{\boldsymbol{P}}
\newcommand{\bH}{\boldsymbol{H}}

\newcommand{\bA}{\boldsymbol{A}}
\newcommand{\pA}{\cb{\cp{A}}}

\newcommand{\bR}{\boldsymbol{R}}
\newcommand{\bS}{\boldsymbol{S}}

\newcommand{\bU}{\boldsymbol{U}}

\newcommand{\bY}{\boldsymbol{Y}}

\newcommand{\bxi}{\boldsymbol{\xi}}

\newcommand{\bSig}{\boldsymbol{\Sigma}}
\newcommand{\bTheta}{\boldsymbol{\Theta}}
\newcommand{\bThetac}{\boldsymbol{\Theta}_\perp}
\newcommand{\bsig}{\boldsymbol{\sigma}}

\newcommand{\bgam}{\boldsymbol{\gamma}}
\newcommand{\bI}{\boldsymbol{I}}

\newcommand{\N}[1]{\cp{N}_{#1}}

\newcommand{\E}{\mathbb{E}}

\newcommand{\tr}{\text{trace}}
\newcommand{\vc}{\text{vec}}
\newcommand{\col}{\text{col}}

\newtheorem{theorem}{Theorem}

\newtheorem{assumption}{Assumption}
\newtheorem{lemma}{Lemma}
\newtheorem{corollary}{Corollary}

\begin{document}

\title{Multitask diffusion adaptation over networks \\ with common latent representations}
\author{Jie Chen, \IEEEmembership{Member, IEEE}, C{\'e}dric Richard, \IEEEmembership{Senior Member, IEEE}, \\
Ali H. Sayed, \IEEEmembership{Fellow Member, IEEE}

\thanks{Copyright (c) 2016 IEEE. Personal use of this material is permitted. However, permission to use this material for any other purposes must be obtained from the IEEE by sending a request to pubs-permissions@ieee.org.}
\thanks{The work of J. Chen was supported in part by the NSFC grant 61671382. The work of C.~Richard was supported in part by the Agence Nationale pour la Recherche, France, (ODISSEE project, ANR-13-ASTR-0030). The work of A. H. Sayed was supported in part by NSF {grants} CIF-1524250 and ECCS-1407712. A short and preliminary version of this work appears in the conference publication~\cite{Chen2014MLSP}. J. Chen is with CIAIC of School of Marine Science and Technology, Northwestern Polytechinical University, Xi'an, 710072, China (e-mail: dr.jie.chen@ieee.org). C. Richard is with the Universit\'e de Nice Sophia-Antipolis, CNRS, France (e-mail: cedric.richard@unice.fr), in collaboration with Morpheme team (INRIA Sophia-Antipolis). A. H. Sayed is with the department of electrical engineering, University of California, Los Angeles, CA 9005-1594, USA (email: sayed@ee.ucla.edu).} }


\maketitle

\vspace{-1cm}
%
%
%

\begin{abstract}

Online learning with streaming data in a distributed and collaborative {manner} can be useful in a wide range of applications. This topic {has been receiving} considerable  attention {in recent years with emphasis on both single-task and multitask scenarios. In single-task adaptation, agents cooperate to track an objective of common interest, while in multitask adaptation agents track multiple {objectives} simultaneously}.  Regularization is {one useful technique to promote and exploit similarity among tasks {in the latter scenario}. This work examines an alternative way to model relations among tasks by assuming that they all share a}  common latent feature representation. {As a result, a new multitask learning formulation is presented and algorithms are developed for its solution in a distributed online manner.}  We present a unified framework {to analyze the mean-square-error performance of the adaptive strategies, and conduct simulations to illustrate the theoretical findings and potential applications.}
\end{abstract}



\vspace{-2mm}
\section{Introduction}

{M}{ulti-agent} networks usually {consist of} a large number of interconnected agents or nodes. Interconnections  {between the agents allow them} to share information and {collaborate} in order {to solve complex tasks collectively}. Examples abound in the realm of social, economic and {biological} networks. Distributed algorithms over {such} networks offer a valuable alternative to centralized solutions with {useful} properties such as scalability, robustness, and decentralization. When endowed with {adaptation} abilities, these algorithms enable agents to continuously learn and adapt in an online {manner} to concept drifts in {their} data streams~\cite{Sayed2014IEEE,Sayed2014adaptation}. {Broadly}, distributed strategies for online parameter estimation can be {applied to single-task or multi-task scenarios}. In the first {case},  agents cooperate with each other to estimate a single parameter vector {of interest, such as tracking a common target}. Reaching consensus among the agents is {critical for successful inference in these problems}. In the {multitask case}, the agents cooperate {to estimate} multiple parameter vectors simultaneously, {such as tracking a collection of targets moving in formation}~\cite{Chen2014multitask}.

Extensive studies {have been} conducted on {adaptive} distributed strategies for single-task problems. Existing techniques include incremental~\cite{Bertsekas1997,Rabbat2005,Blatt2007,Lopes2007incr}, consensus~\cite{Nedic2009,Kar2009,Srivastava2011}, and diffusion strategies{~\cite{Chen2014MLSP,Sayed2014IEEE,Sayed2013diff,Sayed2013intr,Lopes2008diff,Cattivelli2010diff,ChenUCLA2012,ChenUCLA2013}. 
{Incremental techniques require determining a cyclic path that runs across all nodes, which is generally a challenging (NP-hard) task to perform. Besides, feature makes the incremental strategies sensitive to link failures and problematic for adaptation. Consensus techniques aim to reach an agreement among nodes on the estimate of interest via local information exchanges, but they have been shown~\cite{Sayed2014IEEE,Sayed2014adaptation} to suffer from instability problems when used in the context of adaptive networks due to an inherent asymmetry in the update equations.  Diffusion techniques, on the other hand, have been shown to have superior stability and performance ranges~\cite{Tu2012} than consensus-based implementations. For these reasons, we shall focus on diffusion-type implementations in this paper.}

 {Besides single-task scenarios}, there are also applications where it is desirable to estimate multiple parameter vectors at the same time, rather than promote consensus among all agents~\cite{Chen2015diffusion}.} For {example}, geosensor networks that monitor dynamic spatial fields, such as temperature or windspeed {variations} in geographic environments, {require} node-specific estimation problems {that are able to take advantage of the} spatial correlation between the measurements of neighboring nodes~\cite{Abdolee2014,Hirata2008}. {A second example is the problem of} collaborative target tracking where agents track several {objects} simultaneously~\cite{Chen2014multitask,Chen2015diffusion}. {{Motivated by these applications,} there have been several variations of distributed strategies to deal with multitask {scenarios as well.} {Existing strategies mostly depend on how the tasks relate to each other and on exploiting some prior information. In a first scenario, nodes are grouped into clusters, and each cluster of nodes is interested in estimating its own parameter vector. Although clusters may generally have distinct though related estimation tasks to perform, the nodes may still be able to capitalize on inductive transfer between clusters to improve their estimation accuracy. Multitask diffusion strategies were developed to perform estimation under these conditions~\cite{Chen2014multitask,nassif2015multitask}. One useful way to do so is to employ regularization. A couple of other useful works have also addressed variations of this scenario where the only available information is that clusters may exist in the network but nodes do not know which other nodes share the same estimation task~\cite{Zhao2015,Chen2015Eusipco,nassif2016diffusion}.  {In~\cite{Monajemi2016},  the authors use multitask diffusion adaptation with a node clustering strategy to identify a model between the gait information and electroencephalographic signals. In~\cite{Wang2016}, the authors consider the framework in~\cite{Chen2014multitask} to devise a distributed strategy that allows each node in the network to locally adapt inter-cluster cooperation weights. The authors in~\cite{Roula2016} promote cooperation between clusters with $\ell_1$-norm co-regularizers. They derive a closed-form expression of the proximal operator, and introduce a strategy that also allows each node to automatically set its inter-cluster cooperation weights. The works in~\cite{Monajemi2015,Khawatmi2015} propose alternative node clustering strategies.} In a second scenario, it is assumed that there are parameters of global interest to all nodes in the network, {a collection of parameters of common interest within sub-groups of nodes,} and a set of parameters of local interest at each node. A diffusion strategy was developed to perform estimation under these conditions~\cite{Plata2015TSP,Plata2016ICASSP}. Likewise, {in the works~\cite{Bertrand2010P1,Bertrand2011,Bogdanovic2014TSP}}, distributed algorithms are derived to estimate node-specific parameter vectors that lie in a common latent signal subspace. In another work~\cite{Chen2016IcasspGroup}, the diffusion LMS algorithm is extended to deal with structured criteria built upon groups of variables, leading to a flexible framework that can encode various structures in the parameters. An unsupervised strategy to differentially promote or inhibit collaboration between nodes depending on their group is also introduced.}}

{
 Alternatively, in recent years, there has been an increasing interest in modeling relations between tasks by assuming that all tasks share a common feature representation in a latent subspace~\cite{Baxter2000,Ando2005,Zhou2011}. The authors in \cite{Ando2005} proposed a non-convex method based on Alternating Structure Optimization~(ASO) for identifying the task structure. A convex relaxation of this approach was developed in \cite{Chen2009mtl}. In \cite{Zhou2011}, the authors showed the equivalence between ASO,  clustered multitask learning~\cite{Jacob2009,Obozinski2010} and their convex relaxations. The efficiency of such task relationships has been demonstrated in these works for clustering and classification problems. In our preliminary work~\cite{Chen2014MLSP}, we introduced this framework within the context of distributed online adaptation over networks. Useful applications can be envisaged. First, consider the case where the common subspace is spanned by certain selected columns of the identity matrix. This means that a subset of the entries of the parameter vector to be estimated are common to all nodes while no further restriction is imposed on the other entries. Another example concerns beamforming for antenna arrays with a generalized side-lobe canceller (GSC). The latent subspace corresponds to the space where interfering signals reside~\cite{Griffiths1982}. A third example deals with cooperative spectrum sensing in cognitive radios, where the common latent subspace characterizes common interferers~\cite{Plata2015TSP}.}


{Drawing on these motivations}, this paper deals with distributed learning and adaptation over multitask networks with common latent representation subspaces. Algorithms are designed accordingly, and their performance analyzed. The contributions of this work include the following main aspects:
\begin{itemize}
\item We {formulate} a new multitask estimation problem, which assumes that all tasks share a common latent subspace representation in addition to node-specific contributions. Additional constraints can be incorporated if needed. This work contrasts with earlier works~\cite{Chen2014multitask,Roula2016}, where the inductive transfer between learning tasks is promoted by regularizers. It also differs from~\cite{Plata2015TSP}, which considers direct models by stacking local and global variables in an augmented parameter vector. Moreover, the work~\cite{Ando2005}  uses a similar inductive transfer model but the common latent subspace is unknown and embedded into a joint estimation process.  Our work  is the first one to introduce an online estimation algorithm over networks. Estimating the common latent subspace of interest within this context is a challenging perspective.
\item We {explain how this formulation} can be tailored to fit individual application contexts by considering additional model constraints. We illustrate this {fact by considering}  two convex optimization problems and the associated distributed {online} algorithms. The first {algorithm} is a generalization in some sense of the diffusion LMS algorithm, which can be retrieved by defining the low-dimensional common latent subspace as the whole parameter space. The second algorithm uses $\ell_2$-norm regularization to account for the multitask nature of the problem. This opens the way to other regularization schemes depending on the application at hand.
\item We present a unified framework for analyzing the performance of these {algorithms}. This framework also allows to address the performance analysis of the {multitask} algorithms in~\cite{Zhao2012,Chen2014multitask,nassif2014performance,Chen2015diffusion} in a generic {manner}, though these analyses were performed independently of each other in these works.
\end{itemize}


The rest of {the} paper is organized as follows. Section~II introduces the multitask estimation problem considered in this paper. Then, two distributed learning strategies are derived in Section~III by imposing different constraints on common and node-specific representation subspaces. Section~IV provides a general framework for analyzing distributed algorithms of this form. In Section~V, experiments are conducted to illustrate the characteristics of these algorithms. Section~VI concludes the paper and connects our work with several other {learning} {strategies.}

\medskip

\noindent\textbf{Notation.}  Normal font $x$ denotes scalars. Boldface small letters $\bx$ denote vectors. All vectors are column vectors. Boldface capital letters $\cb{X}$ denote matrices. The asterisk $(\cdot)^*$ denotes complex conjugation for scalars and complex-conjugate transposition for matrices. The superscript $(\cdot)^\top$ represents transpose of a matrix or a vector, and $\|\!\cdot\!\|$ is the $\ell_2$-norm of its matrix or vector argument. $\text{Re}\{\cdot\}$ and $\text{Im}\{\cdot\}$ denote the real and imaginary parts of their complex argument, respectively. Matrix trace is denoted by trace$(\cdot)$. The operator $\col\{\cdot\}$ stacks its vector arguments on the top of each other to generate a connected vector. The operator $\text{diag}\{\cdot\}$ formulates a (block) diagonal matrix with its arguments. Identity matrix of size $N\times N$ is denoted by $\bI_N$. Kronecker product is denoted by $\otimes$,  and expectation is denoted by $\E\{\cdot\}$.  We denote by $\N{k}$ the set of node indices in the neighborhood of node~$k$, including $k$ itself, and $|\N{k}|$  its set cardinality.

\vspace{-2mm}
\section{Matched subspace estimation over multitask networks}

\subsection{Multitask estimation problems over networks}

Consider a connected network composed of $N$ nodes. The problem is to estimate an $L\times 1$ unknown vector $\bw_k^o$ at each node $k$ from collected measurements.  {At each time $n$,} node~$k$ has access to local streaming measurements $\{d_k(n), \bx_{k,n}\}$, where $d_k(n)$ is a scalar zero-mean reference signal, and $\bx_{k,n}$ is a $1\times L$ zero-mean row regression vector with covariance matrix $\bR_{x,k}=\E\{\bx_{k,n}^*\bx_{k,n}\} > 0$. The data at agent $k$ and time $n$ are assumed to be related via the linear model:
\begin{equation}
           \label{eq:datamodel}
            d_k(n) = \bx_{k,n}\bw_k^{o} + z_k(n)
\end{equation}
where $\bw_k^o$ is an unknown complex parameter vector, and $z_k(n)$ is a zero-mean i.i.d. noise with variance $\sigma_{z,k}^2 = \E\{|z_k(n)|^2\}$. {The noise signal} $z_k(n)$ is assumed to be independent of any other signal. Let $J_k(\bw)$ be a differentiable convex cost function at agent~$k$. In this paper, we shall consider the mean-square-error criterion:
\begin{equation}
          \label{eq:MSE}
           J_k(\bw) = \mathbb{E} \{ |d_k(n) - \bx_{k,n}\bw|^2\}
\end{equation}
It is clear from~\eqref{eq:datamodel} that each $J_k(\bw)$ is minimized at $\bw_k^o$. {We refer to each parameter $\bw_k^o$ to estimate (or model in a more general sense) as a task. Depending on whether the minima of all $J_k(\bw)$ are achieved at the same $\bw_k^o$ or not,} the distributed learning problem can be single-task or multitask oriented~\cite{Chen2014multitask}.

With single-task networks, all agents aim at estimating the same parameter vector $\bw^o$ shared by the entire network, that is,
\begin{equation}
	\bw_k^o = \bw^o
\end{equation}
for all $k \in \{1, ..., N\}$. Several popular collaborative  strategies, such as diffusion LMS{~\cite{Chen2014MLSP,Sayed2014IEEE,Lopes2008diff,Sayed2013intr}}, were derived to address this problem by seeking the minimizer of the following aggregate cost function:
\begin{equation}
	\label{eq:Jglob}
	J^{\text{glob}}(\bw) = \sum_{k=1}^N J_k(\bw)
\end{equation}
in a distributed manner.  Since the individual costs~\eqref{eq:MSE} admit the same solution, $\bw^o$ is also the solution of~\eqref{eq:Jglob}. {It has been shown that using} proper cooperative strategies to solve~\eqref{eq:Jglob} can improve the estimation performance~\cite{Sayed2014IEEE,Sayed2014adaptation}.

With multitask networks, each agent aims at determining a local parameter vector $\bw_k^o$. It is assumed that some similarities or {relations} exist among the parameter vectors of neighboring agents so that cooperation can still be meaningful, namely,
\begin{equation}
	\bw_k^o \sim \bw_\ell^o \, \text{ if } \ell \in \N{k}
\end{equation}
where {the} symbol $\sim$ {refers to} a similarity relationship in some sense, which can be exploited to enhance  performance. Depending on {the} problem characteristics, this property can be promoted {in several ways}, e.g., by introducing some regularization term, {or by assuming} a common latent structure.  Networks may also be structured into clusters where agents within each cluster estimate the same parameter vector{~\cite{Chen2014multitask,Zhao2012}}.

\vspace{-2mm}
\subsection{Node-specific subspace constraints}

Although agents aim to estimate distinct minimizers $\bw_k^o$, exploiting relationships between solutions can make cooperation among agents beneficial. Regularization is {one} popular technique for introducing prior information about the solution. It can improve estimation accuracy though it may introduce {bias}~\cite{Chen2015diffusion,Chen2014multitask,Chen2013performance}. In this paper, we explore an alternative strategy that assumes that the hypothesis spaces partially overlap. {Specifically, we assume that each {$\bw_k^o$} can be expressed in the form:}
\begin{equation}
	\label{eq:structure}
	\bw_k^o = \bTheta\bu^o + \beps_k^o
\end{equation}
where $\bTheta\bu^o$ is common to all nodes with $\bTheta$ {denoting} an $L\times M$ matrix with known entries {and $\cb{u}^o$ an unknown $M \times 1$ parameter vector (common to all nodes)},  and where $\beps_k^o$  {is} an unknown node-specific component. We assume that matrix $\bTheta=[\btheta_1,\dots, \btheta_M]$ is full-rank with $M \leq L$. Overcomplete sets of column vectors $\{\btheta_1,\dots, \btheta_M\}$ may be {advantageous in some} scenarios but {this} usually requires to impose further constraints such as sparsity over $\bu^o$. We shall not discuss this case further in order to focus on the main points of the presentation. Model~\eqref{eq:structure} means that all tasks share the same parameter vector $\bTheta\bu^o${, which} lies in the subspace spanned by columns of $\bTheta$. 
This subspace representation can be useful in several applications.  For instance, consider the case where  $\bTheta$ is composed of selected columns of the identity matrix $\bI_L$. This means that a subset of the entries of $\bw^o_k$ are common to all agents while no further assumptions are imposed on the other entries. This situation is {a natural} generalization of the single-task scenario. Another example concerns beamforming {problems with a} generalized sidelobe canceller (GSC), where $\bTheta$ acts as a blocking matrix to cancel signal components that lie in the constraint space~\cite{Griffiths1982}. In machine learning, formulation \eqref{eq:structure} is referred to as the alternating structure optimization~(ASO) problem~\cite{Ando2005,Zhou2011}. The subspace $\bTheta$ is, however, learnt simultaneously via {a} non-convex optimization procedure.  In what follows, we shall assume that $\bTheta$ is known by each agent. 

{Before proceeding further, we clarify the difference between model~\eqref{eq:structure} addressed here and in our preliminary work~\cite{Chen2014MLSP}, and the model studied in~\cite{Bogdanovic2014TSP,Plata2013,Plata2016ICASSP,Bogdanovic2014,Plata2015TSP}. 
In these last works, the authors consider particular information access models where global and local components are assumed to be related to distinct regressors. The centralized problem can then be formulated by stacking the global and local regressors, and by considering a parameter vector augmented accordingly. In our work, motivated by applications of the latent space model in batch-mode learning, we address the problem where the parameter vectors to be estimated lie in global and local latent subspaces. We do not need to distinguish explicitly between global and local regressors. Instead, as shown in the sequel, some extra conditions are needed so that model~\eqref{eq:structure} is identifiable. Among other possibilities, we shall investigate two strategies where constraints on $\bTheta$ and $\beps_k^o$ are imposed.}

Replacing~\eqref{eq:structure} into~\eqref{eq:MSE}, the global cost function is expressed as a function of {a} common {parameter} $\bu$ and node-specific perturbations $\{\beps_k\}_{k=1}^N$:
\begin{equation}
	\label{eq:Jglob.str}
	J^\text{glob}\, \big(\bu, \{\beps_k\}_{k=1}^N\big) 
	= \sum_{k=1}^N \E\big\{|d_k(n) - \bx_{k,n}(\bTheta\bu+\beps_k)|^2\big\}
\end{equation}
We expect the estimation of {$\bw_k^o$}  by each agent to benefit from the cooperative estimation of $\bu$. {Problem~\eqref{eq:Jglob.str} is still insufficient for estimating the tasks $\{\bw_k^o\}$. This is because} the decomposition $\bw_k = \bTheta\, \bu+\beps_k$ is not unique. Indeed, given any optimum solution $\{\bar{\bu}, \bar{\beps}_k\}$, {and any $\cb{s} = \bTheta\bx$, we can generate another optimum solution by  considering the shift $\{\bar{\bu}-\cb{x}, \bar{\beps}_k+\cb{s}\}$.} This ambiguity prevents us from deriving collaboration strategies based on $\bu$. From the point of view of convex analysis, the Hessian matrix of~\eqref{eq:Jglob.str} is rank deficient and no unique solution exists.

\vspace{-2mm}
\section{Problem formulations and solution algorithms}

Problem~\eqref{eq:Jglob.str} can be modified to make it well-determined and more meaningful. In this section, among other possibilities, we investigate two strategies that consist of imposing further constraints and derive the corresponding distributed algorithms. These two formulations guarantee the uniqueness of the solution and have clear interpretations.

\vspace{-4mm}
\subsection{Node-specific subspace constraints}

We restrict the node-specific components $\{\beps_k\}_{k=1}^N$ to lie in the complementary subspace to $\text{span}(\bTheta)$. The problem can be formulated as:
\begin{equation}
	\label{eq:P1}
	\begin{split}
	\min_{\bu,  \{\beps_k\}_{k=1}^N}& J^\text{glob}\big(\bu, \{\beps_k\}_{k=1}^N\big)  \\
	&\text{ subject to } \beps_k\in \text{span}(\bThetac), \quad \forall k = 1, \dots, N
	\end{split}
\end{equation}
where the $L-M$ columns of matrix $\bThetac$ span the complementary subspace to $\text{span}(\bTheta)$, that is, $\bTheta^* \, \bThetac = \cb{0}$. We write:
\begin{equation}
	\label{eq:eps_xi}
	\beps_k = \bThetac\,\bxi_k
\end{equation}
where $\bxi_k$ is a column vector of size $(L-M)$. Now,  replacing~\eqref{eq:eps_xi} into~\eqref{eq:P1}, the  optimization problem becomes unconstrained and the objective function is given by:
\begin{align}
	\label{eq:Jglob.perp}
		&J^\text{glob}\big(\bu, \{\bxi_k\}_{k=1}^N\big)  \nonumber \\
		&= \sum_{k=1}^N \E\big\{|d_k(n) - \bx_{k,n}(\bTheta\bu+\bThetac\bxi_k)|^2\big\}  \nonumber\\
		&= \sum_{k=1}^N \E\{|d_k(n)|^{2}\} + \bu^*\bTheta^*  {\Big(} \sum_{k=1}^N \bR_{x,k} {\Big)}\bTheta\bu		
		+ \sum_{k=1}^N  \bxi_k^*\bThetac^*  \bR_{x,k}\bThetac\bxi_k +2\, \text{Re}\Big\{  \bu^*\bTheta^* \sum_{k=1}^N\bR_{x,k}\bThetac\bxi_k \Big\} \nonumber\\
		&-2\, \text{Re}\Big\{  \sum_{k=1}^N \bp_{dx,k}^*\bTheta\bu \Big\}
		-2\, \text{Re}\Big\{\sum_{k=1}^N \bp_{dx,k}^*\bThetac\bxi_k\Big\}
\end{align}
{where $\bR_{x,k}=\E\{\bx_{k,n}^*\bx_{k,n}^{\phantom{*}}\}$ is the covariance matrix of $\bx_{k,n}$, and $\bp_{dx,k}=\E\{d_{k}(n)\bx^*_{k,n}\}$} is the covariance vector between {the} input data $\bx_{k,n}$ and {the} reference output data $d_k(n)$.

\lemma Problem~\eqref{eq:P1} has a unique solution with respect to $\bu$ and  $\{\beps_k\}_{k=1}^N$ if the perturbations $\{\beps_k\}_{k=1}^N$ lie in a subspace orthogonal to $\text{span}(\bTheta)$.  \endproof

{Proof of Lemma 1 is provided in Appendix~\ref{app:proof-lemma1}.} We shall now derive a distributed algorithm to {seek the minimizer of}~\eqref{eq:P1}. Focusing on the terms that depend on $\bu$ in~\eqref{eq:Jglob.perp}, and setting parameters $\bxi_k$ to their optimum values $\bxi_k^o$, {we} consider first the global cost function {over} the variable $\bu$:
\begin{align}
	J_u^\text{glob}(\bu) &=   \sum_{k=1}^N \Big( \bu^*\bTheta^* \bR_{x,k}\bTheta\bu 
	+ 2 \, \text{Re}{\Big\{} \bu^*\bTheta^*\bR_{x,k}\bThetac\bxi_k^o  {\Big\}} 
	 - 2 \, \text{Re}\Big\{ \bp_{dx,k}^*\bTheta\bu\Big\}
	+ g_k(\bxi_k^o) \Big)  \nonumber \\
	& =   \sum_{k=1}^N J_{u,k} (\bu) 
\end{align}
where  $g_k(\bxi_k^o)$ collects all the terms depending only on $\bxi_k^o$ in~\eqref{eq:Jglob.perp}. {The term $\sum_{k=1}^N \E\{|d_k(n)|^{2}\}$ is discarded because it is constant with respect to the arguments $\bu$ and $\{\bxi_k\}_{k=1}^N$.} Since $J_u^\text{glob}(\bu) $ has a unique minimizer for all nodes over the network, we can use a single-task {adapt-then-combine (ATC) diffusion} strategy to estimate $\bu^o$~\cite{Sayed2013intr,Cattivelli2010diff}. {We introduce a right-stochastic matrix $\bC$ with nonnegative entries $c_{\ell k}$ such that:
\begin{equation} 
	\begin{array}{lcl}\displaystyle
	\sum_{k=1}^N c_{\ell  k} =1,  	&\text{ and }&  c_{\ell k} = 0 \text{ if } k \notin \N{\ell} \\
	\end{array}
\end{equation}
With each node $k$, we associate the local cost over the variable~$\bu$:
\begin{equation}
	J_{u,k}^{\text{loc}} (\bu) = \sum_{\ell\in\N{k}} c_{\ell k} J_{u,\ell} (\bu)
\end{equation}
Observe that $\sum_{k=1}^N J_{u,k}^{\text{loc}} (\bu) =  J_u^\text{glob}(\bu) $ because matrix $\bC$ is right-stochastic. Since $J_u^\text{glob}(\bu)$ is quadratic with respect to $\bu$, it can be expressed at each node $k$ as follows:
\begin{equation}
	\label{eq:Juglob2}
	\begin{split}
         J_u^\text{glob}(\bu) 
         	&= J_{u,k}^{\text{loc}} (\bu)  + \sum_{\ell \neq k}J_{\bu,\ell}^{\text{loc}} (\bu)  \\
		&= J_{u,k}^{\text{loc}} (\bu)  + \sum_{\ell \neq k} \|\bu-\bu^o\|^2_{\nabla^2 J_{u,\ell}^{\text{loc}}}
	\end{split}
\end{equation}
where $\nabla^2 J_{u,\ell}^{\text{loc}}$ denotes the Hessian matrix of $J_{u,\ell}^{\text{loc}}(\bu)$ with respect to $\bu$, and $\|\bu\|^2_{\bSig}$ is the squared norm of $\bu$ weighted by any positive semi-definite matrix $\bSig$, i.e., $\|\bu\|^2_{\bSig} = \bu^* \bSig \bu$. Following an argument based on the Rayleigh-Ritz characterization of eigenvalues~\cite[Sec. 3.1]{Sayed2013intr}, we approximate ${\nabla^2 J_{\bu,\ell}^{\text{loc}}}$  by a multiple of the identity matrix, so that $\|\bu-\bu^o\|^2_{\nabla^2 J_{\bu,\ell}^{\text{loc}}} \approx b_{\ell k}  \|\bu-\bu^o\|^2$.}

{Minimizing~\eqref{eq:Juglob2} in two successive steps yields: 
\begin{align}
               {\bphi}_{k,n} &= \bu_{k,n-1}  - \mu \nabla J_{u,k}^{\text{loc}} (\bu_{k,n-1})  \label{eq:uop1} \\
               \bu_{k,n} &=  {\bphi}_{k,n}  + \mu \sum_{\ell \neq k} b_{\ell k}(\bu^o - \bu_{k,n-1})    \label{eq:uop2}
\end{align}
where {$\mu$ is a positive step size. Its choice to ensure stability of the algorithm will be elaborated on later in Sec. IV}. Now, note the following. First, iteration~\eqref{eq:uop2} requires knowledge of $\bu^o$, which is not available. Each node $\ell$ has a readily available, however, an approximation for $\bu^o$, which is ${\bphi}_{k,n}$. Therefore, we replace $\bu^o$ by ${\bphi}_{k,n}$ in~\eqref{eq:uop2}. Second, ${\bphi}_{k,n}$ at node $k$ is generally a better estimate for $\bu^o$ than $\bu_{k,n-1}$ since it is obtained by incorporating information from the neighbors through \eqref{eq:uop1}. Therefore, we replace $\bu_{k,n-1}$ by ${\bphi}_{k,n}$ in \eqref{eq:uop2}. Then, absorbing coefficients $b_{\ell k}$ into another set of nonnegative coefficients that satisfies:
\begin{equation}
	\begin{array}{lcl} \displaystyle
	\sum_{\ell=1}^N a_{\ell  k} =1,  	&\text{ and }&  a_{\ell k} = 0 \text{ if } \ell \notin \N{k}, 
	\end{array}
\end{equation} 
which means that matrix $\bA$ with entries $a_{\ell k}$ is left-stochastic, using an instantaneous approximation of the gradient, and  limiting the summation in \eqref{eq:uop2} to the neighbors of node~$\ell$ (see~\cite{Sayed2013intr,Cattivelli2010diff} for more details on a similar derivation in the context of single-task diffusion strategies), we can update $\bu_{k,n}$ as follows:
}

\begin{align}
	\label{eq:upu-1}
	{\bphi}_{k,n} &= \bu_{k,n-1} + \mu \, \sum_{\ell\in\N{k}} c_{\ell k}
	\bTheta^*\bx_{\ell,n}^* \big[d_{\ell}(n) - \bx_{\ell,n}(\bTheta\bu_{k,n-1}) 
	- \bx_{\ell,n} (\bThetac\bxi_{\ell,n-1})\big] \\
	\label{eq:upu-2}
	\bu_{k,n} &= \sum_{\ell\in\N{k}} a_{\ell k}   {\bphi}_{\ell,n}
\end{align}
where $\bxi_{k,n-1}$ is an estimate for the unknown minimizer $\bxi_k^o$, to be evaluated as explained further ahead in~\eqref{eq:upxi}.

Focusing on the terms that depend on $\{\bxi_k\}_{k=1}^N$ in~\eqref{eq:Jglob.perp}, and setting parameter $\bu$ to its optimum value $\bu^o$, {we} consider the global cost function {over} the variables $\bxi_k$:
\begin{equation}
	\begin{split}
	&J_\xi^\text{glob}\big(\{\bxi_k\}_{k=1}^N\big)  \\
		& = 	\sum_{k=1}^N  \Big(\bxi_k^*\bThetac^*\bR_{x,k}\bThetac\bxi_k
			+2\,\text{Re}\Big\{  \bxi_k^*{\bThetac^*\bR_{x,k}\bTheta\bu^o}  \Big\} 
			- 2\,\text{Re}\Big\{\bp_{dx,k}^*\bThetac\bxi_k\Big\}\Big) + g'_k(\bu^o) \\
		& =   \sum_{k=1}^N J_{\xi,k} (\bxi_k) 
	\end{split}
\end{equation}
where  $g'_k(\bxi_k^o)$ collects all the terms depending only on $\bu^o$ in~\eqref{eq:Jglob.perp}. Now since the parameters $\bxi_{k}$ are node-specific, if no further constraints are imposed, they can be updated independently of each other via an LMS-type update:
\begin{equation}
	\label{eq:upxi}
	\begin{split}
	&\bxi_{k,n} = \bxi_{k,n-1} +\mu \bThetac^*\bx_{k,n}^* \big[d_{k}(n) 
	- \bx_{k,n}(\bTheta\bu_{k,n-1} {+}\,\bThetac\bxi_{k,n-1})\big]   
	\end{split}
\end{equation}
At each time instant $n$, node $k$ updates its parameters $\bu_{k,n-1}$ and $\bxi_{k,n-1}$ using~\eqref{eq:upu-1}--\eqref{eq:upu-2} and~\eqref{eq:upxi}, respectively. The local estimate $\bw_{k,n}$ is {then} given by:
\begin{equation}
         \label{eq:upw}
          \bw_{k,n} = \bTheta\bu_{k,n} + \bThetac\bxi_{k,n}
\end{equation}
It is interesting to {note} that we can rewrite the algorithm without using the auxiliary variables $\bu_{k{,n}}$ and $\{\bxi_{k{,n}}\}_{k=1}^N$, by {substituting} the relations:
\begin{align}
	\label{eq:u-proj}
	\bu_{k{,n}} &=(\bTheta^*\bTheta)^{-1}\bTheta^*\bw_{k{,n}}  \\
	\label{eq:xi-proj}
	\bxi_{k{,n}} &= (\bThetac^*\bThetac)^{-1}\bThetac^*\bw_{k{,n}}
\end{align}
into~\eqref{eq:upu-1}--\eqref{eq:upu-2} and~\eqref{eq:upxi}, respectively. {Selecting} $\bC=\bI_N$ to avoid exchanging raw data and node-specific components, we can implement the update of $\bw_{k,n-1}$ to an intermediate value $\bpsi_{k,n}$ as follows:
\begin{align}
           \bpsi_{k,n} 
		&\stackrel{(a)}{=} \bTheta \bphi_{k,n} + \bThetac \bxi_{k,n}  \nonumber \\
		&\stackrel{(b)}{=} \bTheta \bu_{k,n-1} + \bThetac \bxi_{k,n-1} + \mu \Big[ (\bTheta \bTheta^*  
           		+ \bThetac \bThetac^*) 
	                 \bx_{k,n}^* [ d_{k}(n) 
			- \bx_{k,n}(\bTheta\bu_{k,n-1}+\bThetac\bxi_{k,n-1})] \Big]  \nonumber \\
		& = \bw_{k,n-1} + \mu \bS_{\bTheta}\bx_{k,n}^*\big[d_{k}(n) -\bx_{k,n}\bw_{k,n-1}\big]
\end{align}           
with $\bS_{\bTheta}=\bTheta\bTheta^* + \bThetac^{\phantom{\top}}\bThetac^*$. For step (a), we use \eqref{eq:upw} with the intermediate value $\bphi_{k,n}$ of $\bu_{k,n}$ in \eqref{eq:upu-1} and $\bxi_{k,n}$. Step~(b) follows from their adaptation steps \eqref{eq:upu-1} and \eqref{eq:upxi}.
Now substituting~\eqref{eq:upu-2} in~\eqref{eq:upw} to aggregate the intermediate estimates of $\bu_{k,n}$ from the neighbors of node $k$, we arrive at the combination step:
\begin{align}
              \bw_{k,n} &\stackrel{\eqref{eq:upw}}{=} \bTheta \bu_{k,n} +  \bThetac \bxi_{k,n}  \nonumber \\
                              & \stackrel{(c)}{=} \bTheta \!\!\!   \sum_{\ell\in\N{k}}\!\!a_{\ell k} (\bTheta^*\bTheta)^{-1}\!\bTheta^* \!\bpsi_{k,n} 
                              \!\!+\! \bThetac  (\bThetac^*\bThetac)^{-1}\, \bThetac^* \bpsi_{k,n}  \nonumber \\
                              & =  \sum_{\ell \in \N{k}} a_{\ell k}  \bP_{\bTheta}  \bpsi_{\ell,n} + \bP_{\bThetac} \bpsi_{k,n} 
\end{align}
where $\bP_{\bTheta}=\bTheta(\bTheta^*\bTheta)^{-1} \bTheta^*$ and $\bP_{\bThetac}=\bI_{{L}} - \bP_{\bTheta}$ {are} the projection matrices over subspaces $\bTheta$ and $\bThetac$. For step (c), we use \eqref{eq:u-proj}--\eqref{eq:xi-proj} with the intermediate estimate $\bpsi_{k,n}$. 
Finally, we arrive at the ATC strategy summarized in Algorithm~\ref{algo:LMSmult1}.

The first step in \eqref{eq:Alg1_1} is an adaptation step where node $k$ uses the data realizations $\{d_k(n),\bx_{k,n}\}$ to update its existing estimate $\bw_{k,n-1}$ to an intermediate value $\bpsi_{k,n}$. All other nodes in the network are performing a similar step. The second step in \eqref{eq:Alg1_2} is an aggregation step. To update its intermediate estimate to $\bw_{k,n}$, each node $k$ combines the existing estimates of its neighbors in the common latent subspace $\bTheta$ to build up a common representation, and refines it with a node-specific value in $\bThetac$. In the special case when $\bA=\bI_N$, so that no information exchange is performed, the ATC strategy reduces to a non-cooperative solution where each node $k$ runs its own individual descent algorithm.

Matrix $\bS_{\bTheta}$ in the adaptation step~\eqref{eq:Alg1_1} is positive-definite. It arises from the calculation of the gradient of~\eqref{eq:Jglob.perp} with respect to $\bu$ and $\bxi_k$. The algorithm can be simplified by replacing $\bS_{\bTheta}$ by $\bI_L$ in~\eqref{eq:Alg1_1} without compromising the convergence of the method (as analyzed further ahead in Section~\ref{sec:analyses}). We then arrive at the recursion:
\begin{equation}
	\label{eq:lms-update}
	\bpsi_{k,n} = \bw_{k,n-1} + \mu\,\bx_{k,n}^*\big[d_{k}(n) -\bx_{k,n}\bw_{k,n-1}\big]
\end{equation}
Strictly speaking, observe that $\bS_{\bTheta}=\bI_L$ if, and only if, the columns of $\bTheta$ and $\bThetac$ form an orthonormal basis of $\R^L$. Note that the adaptation step \eqref{eq:lms-update} is the LMS solution for minimizing the cost in~\eqref{eq:Jglob.perp} with respect to $\bw_k$. 

{Before leaving this section, we would like to point out that the algorithm described in~\cite{Plata2015TSP}, which addresses direct models by stacking global and local variables in an augmented parameter vector, may be used to solve problem~\eqref{eq:P1}, provided that an appropriate variable change is performed in order to make the latent variables $\bu_{k,n}$ and $\bxi_{k,n}$ explicit in $\bw_{k,n}$. The resulting algorithm has the same performance as Algorithm~\ref{algo:LMSmult1} defined by~\eqref{eq:Alg1_1},~\eqref{eq:Alg1_2}, but, obviously, they do not have the same form since they do not operate in the same domain. This structural difference has a major consequence for Algorithm~1. As already explained, it can be further tuned by replacing the matrix $\cb{S}_{\cb{\Theta}}$ in~\eqref{eq:Alg1_1} by any positive definite matrix while ensuring convergence of the method. This extra degree of freedom will be taken into account in the analysis of the algorithm, where the only condition on $\cb{S}_{\cb{\Theta}}$ is to be positive definite. We will also show that setting $\cb{S}_{\cb{\Theta}}$ to $\bI_L$, besides simplifying Algorithm~\ref{algo:LMSmult1}, can greatly improve its performance.
 }

\begin{algorithm}[!t]
\textbf{Parameters:} Preset \\
 \hspace{5ex} - positive step-size $\mu$ for all agents; \\
 \hspace{5ex} - left-stochastic combination matrix $\bA$;\\
 \hspace{5ex} - {full-rank matrix} $\bTheta$ with {columns} $\{\btheta_1, \dots, \btheta_M\}$.
 
\textbf{Initialization:} Set initial weights $\bw_{k,0} = \cb{0}$  for all $k\in\{1, ..., N\}$.
\newline\textbf{Algorithm:}\hspace{2ex} At each time instant $n \geq 1$, and for each agent $k$, update $\bw_{k,n}$ as:
\begin{align}
               \bpsi_{k,n} & = \bw_{k,n-1} \!+\! \mu\bS_{\bTheta}\,\bx_{k,n}^*\big[d_{k}(n) -\bx_{k,n}\bw_{k,n-1}\big]
               \label{eq:Alg1_1}  \\
               \bw_{k,n} &= \sum_{\ell \in \N{k}} a_{\ell k}  \bP_{\bTheta}  \bpsi_{\ell,n} 
               + \bP_{\bThetac} \bpsi_{k,n} 
               \label{eq:Alg1_2}
\end{align}

\caption{ATC diffusion LMS with node-specific subspace constraints \protect{}} \label{algo:LMSmult1}
\vspace{-2mm}
\end{algorithm}

\subsection{Node-specific subspace constraints with norm-bounded projections}

{The second formulation we consider is to} relax the constraint that node-specific components $\{\beps_k\}_{k=1}^N$ must lie in $\text{span}(\bTheta_\perp)$. {We now} assume that they are norm-bounded in some sense. The problem is formulated as follows:
\begin{equation}
	\label{eq:P2}
		\begin{split}
		&\min_{\bu,\{\beps_k\}_{k=1}^N}  J^\text{glob}\big(\bu, \{\beps_k\}_{k=1}^N\big)  \\
		&\text{subject to } \sum_{k=1}^N\|\bP_{\bTheta}\beps_k\|^2 \leq \nu_{1}, \quad
		 \sum_{k=1}^N\|\bP_{\bThetac}\beps_k\|^2 \leq \nu_{2}
              \end{split}
\end{equation}
Since the objective function and the constraints are convex in {$(\bu, \{\beps_k\}_{k=1}^N)$}, the constrained problem \eqref{eq:P2} can be formulated as a regularized optimization problem that consists of minimizing {a global cost of the form}~\cite{boyd2004convex}:
\begin{equation}
            \label{eq:Jglob.pert.reg}
            \begin{split}
            J^\text{glob}\big(\bu, \{\beps_k\}_{k=1}^N\big) &= \sum_{k=1}^N \E\big\{|d_{k}(n) - 
            {\bx_{k,n}(\bTheta\bu+\beps_k)}|^2\big\}+ \eta_1\sum_{k=1}^N \|\bP_{\bTheta}\beps_k\|^2 
            + \eta_2\sum_{k=1}^N \|\bP_{\bThetac}\beps_k\|^2
            \end{split}
\end{equation}
where $\eta_1$ and $\eta_2$ are positive regularization parameters that are related to the bounds $\nu_{1}$ and $\nu_{2}$.

\lemma  Problem~\eqref{eq:P2} has a unique solution with respect to $\bu$ and  $\{\beps_k\}_{k=1}^N$. \endproof

{Proof of Lemma~2 is provided in Appendix~\ref{app:proof-lemma2}.} Other norms such as the general $\ell_{p,q}$-norm may be used with $\beps_k$ in \eqref{eq:P2}, depending on the application. Some form of regularization {on} $\bu$ may also be included. However, using the $\ell_2$-norm with $\beps_k$ in \eqref{eq:P2} enables us to solve the problem with respect to $\bw_k$, without using the auxiliary variables $\bu$ and $\{\beps_k\}_{k=1}^N$. Indeed, let us rewrite~\eqref{eq:Jglob.pert.reg} as follows:
 \begin{equation}
	\label{eq:Jglob.pert.reg2}
	\begin{split}
	&J^\text{glob}\, \big(\bu, \{\bw_k\}_{k=1}^N\big) 
	= \sum_{k=1}^N \E\big\{|d_{k}(n) - \bx_{k,n}\bw_k  |^2\big\} 
	+ \eta_1\sum_{k=1}^N \|\bP_{\bTheta}(\bw_k - \bTheta\bu)\|^2 
	+  {\eta_2 \sum_{k=1}^N \|\bP_{\bThetac}\bw_k\|^2}
	\end{split}
\end{equation}
The optimality {condition relative to $\bu$ gives}:
\begin{equation}
             \sum_{k=1}^N \bTheta^* \bP_{\bTheta} (\bw_k^o - \bTheta\bu^o) = \cb{0}
\end{equation}
from\! which\! the\! optimal\! parameter\! vector\! $\bu^o$\! can\! be\! expressed\! as:
\begin{equation}
             \label{eq:u.opt}
              \bu^o =   \frac{1}{N} \sum_{k=1}^N (\bTheta^*\bTheta)^{-1}\bTheta^* \bw_k^o
\end{equation}
Substituting~\eqref{eq:u.opt} into~\eqref{eq:Jglob.pert.reg2}, and using that $\bP_{\bTheta}$ is Hermitian and idempotent {(i.e., $\bP_{\bTheta} = \bP_{\bTheta}^2$)}, yields:
 \begin{equation}
	\label{eq:Jglob.pert.reg3}
	\begin{split}
	J^\text{glob}\, \big(\{\bw_k\}_{k=1}^N\big) = 
	\sum_{k=1}^N \E\big\{|d_{k}(n) - \bx_{k,n}\bw_k |^2\big\} 
	+ \eta_1\sum_{k=1}^N \Big\| \bP_{\bTheta} \bw_k - \frac{1}{N} \sum_{\ell=1}^N \bP_{\bTheta}\bw_\ell\Big\|^2  
	+ \eta_2\sum_{k=1}^N \|\bP_{\bThetac} \bw_k\|^2
	\end{split}
\end{equation}
Node $k$ can apply a steepest-descent iteration to minimize the cost in~\eqref{eq:Jglob.pert.reg3} {with respect to $\{\bw_k\}_{k=1}^N$}. Computing the gradient vector of~\eqref{eq:Jglob.pert.reg3}  we get:
\begin{align}
	\label{eq:Jglob.pert.grad}
	\nabla J^{\text{glob}} &= \Big[(\bR_{x,k}\bw_k - \bp_{dx})
	+ \eta_1 \big(\bP_{\bTheta} \bw_k -  \frac{1}{N}  \sum_{\ell=1}^N \bP_{\bTheta} \bw_\ell\big) 
	+\eta_2 \bP_{\bThetac} \bw_k\Big]^*
\end{align}
{Starting from an initial condition $\bw_{k,0}$, we arrive at the steepest descent iteration:
\begin{align}
	\bw_{k,n} &= \bw_{k,n-1} - \mu \Big[(\bR_{x,k}\bw_{k,n-1} - \bp_{dx}) +\eta_2 \bP_{\bThetac} \bw_{k,n-1} \Big]  \nonumber \\
	&-\mu \eta_1\Big(\bP_{\bTheta} \bw_{k,n-1} - \frac{1}{N}  \sum_{\ell=1}^N \bP_{\bTheta} \bw_{\ell,n-1}\Big) 
\end{align}
This iteration indicates that the update term involves adding two correction terms to $\bw_{k,n-1}$. Among many other forms, we can implement the update in two successive steps by adding one correction term at a time:
\begin{align}
	\label{eq:diff2prot}
	\bpsi_{k,n} &= \bw_{k,n-1}
	 - \mu\Big[(\bR_{x,k}\bw_{k,n-1} - \bp_{dx}) + \eta_2\bP_{\bThetac}\bw_{k,n-1}\Big]\\
	\label{eq:diff2prot-2}
	\bw_{k,n}   &=   \bpsi_{k,n}  
	- \mu \eta_1\Big(\bP_{\bTheta} \bw_{k,n-1} - \frac{1}{N}  \sum_{\ell=1}^N \bP_{\bTheta} \bw_{\ell,n-1}\Big) 
\end{align}
Step~\eqref{eq:diff2prot} updates $\bw_{k,n-1}$ to an intermediate value $\bpsi_{k,n}$. We now revise~\eqref{eq:diff2prot}--\eqref{eq:diff2prot-2}  to achieve a diffusion LMS {type} algorithm. The intermediate value $\bpsi_{\ell,n}$ at node $\ell$ is {generally expected to be} a better estimate for $\bw_\ell^o$ than $\bw_{\ell,n-1}$ since it is updated by the first step~\eqref{eq:diff2prot}. Therefore, we replace $\bw_{\ell,n-1}$ by $\bpsi_{\ell,n}$ in the second step~\eqref{eq:diff2prot-2} as follows {to get}:}
\begin{equation}
	\begin{split}
	\bw_{k,n}   
	&=   \bpsi_{k,n}  - \mu \eta_1\Big(\bP_{\bTheta} \bpsi_{k,n} - \frac{1}{N}  \sum_{\ell=1}^N \bP_{\bTheta} \bpsi_{\ell,n}\Big)  \label{eq:alg2ao}  \\
	& = \!(\bpsi_{k,n} \!\!-\! \bP_{\Theta} \bpsi_{k,n}) \!+\! \Big(\!(1\!-\!\mu\eta_1)  \bP_{\Theta} \bpsi_{k,n} \!{+}\!\! \sum_{\ell=1}^N\frac{\mu\eta_1}{N}  \,\bpsi_{\ell,n}\!\Big)
        \end{split}
\end{equation}
Observe that $\bP_{\bThetac} \bpsi_{k,n}=\bpsi_{k,n} - \bP_{\bTheta} \,\bpsi_{k,n}$, and introduce the coefficients $a_{\ell k} = \frac{\mu\eta_1}{N}$ for $\ell \neq k $, and $a_{k k} = 1-\mu\eta_1+\frac{\mu\eta_1}{N}$. We get:
\begin{equation}
          \bw_{k,n}   =  \bP_{\bThetac} \bpsi_{k,n}   + \sum_{\ell=1}^{N} a_{\ell k} {\bP_{\bTheta}}\bpsi_{\ell,n}
\end{equation}
Considering that each node in the network can only share information with its neighbors, and using instantaneous approximations for $\bR_{x,k}$ and $\bp_{dx}$, we arrive at:
\begin{align}
	\bpsi_{k,n} 	&= (\bI_L - \mu \eta_2\bP_{\bThetac})\bw_{k,n-1}
				+   \mu\bx_{k,n}^* \big[d_k(n) - \bx_{k,n}\bw_{k,n-1}\big] \\
	\bw_{k,n}   	&= \sum_{\ell\in\N{k}} a_{\ell k} \bP_{\bTheta}  \bpsi_{\ell,n} + \bP_{\bThetac}\bpsi_{k,n}
\end{align}
with $a_{k k} = 1-\mu\eta_1+\frac{\mu\eta_1}{|\N{k}|}$ and $a_{\ell k} = \frac{\mu\eta_1}{|\N{k}|}$ for $\ell \in \N{k}$ and $\ell \neq k$. Note that, for sufficiently small step-sizes $\mu_k$, these coefficients are nonnegative and satisfy $\sum_{\ell=1}^Na_{\ell k}=1$ for all $k$. We will treat these coefficients as free parameters that can be chosen by the designer according to these conditions {(i.e., nonnegative coefficients that add up to one on each column of matrix $\bA$)}. We summarize this {statement} in Algorithm 2.

{Algorithms 1 and 2 employ the same aggregation step in~\eqref{eq:Alg1_2} and~\eqref{eq:Alg2.b}. Node $k$ combines the intermediate estimates of its neighbors in the common subspace $\bTheta$ without affecting the local contribution in the complementary subspace $\bThetac$. The norm constraint \eqref{eq:P2} in $\bThetac$ leads to a leaky-LMS alike term in the adaptation step \eqref{eq:Alg2.a}.}

Let us now examine two special cases of Algorithm 2. First, in the case where $\bTheta = \cb{0}$, problem \eqref{eq:Jglob.pert.reg} reduces to a regularized least-mean {squares} problem with $\bw_k=\beps_k$. {That is, the} algorithm reduces to the non-cooperative leaky-LMS  algorithm. {On the other hand,} if $\bTheta=\bI_L$, the algorithm {reduces} to diffusion LMS.

{Before leaving this section, we briefly discuss the complexity of Algorithms 1 and 2. Both algorithms have the same adapt-then-combine structure as the diffusion LMS except that each node needs to project data on $\boldsymbol{\Theta}$ and $\boldsymbol{\Theta}_\perp$. This means that each node $k$ only needs to update the $L \times 1$ parameter vectors $\boldsymbol{\psi}_{k,n}$ and $\boldsymbol{w}_{k,n}$ at each time instant. Next, each node $k$ needs to transmit $\boldsymbol{w}_{k,n}$ to its $|\cp{N}_k|-1$ neighbors. A projection performed by a matrix-vector product has a computational complexity of $\cp{O}(L \log_2 L)$~\cite{Gohberg1994}. All the other operations performed by each node have a complexity of $\cp{O}(L)$.}

\begin{algorithm}[!t]
\textbf{Parameters:} Preset \\
 \hspace{5ex} - positive step-size $\mu$ for all agents; \\
 \hspace{5ex} - {full-rank matrix} $\bTheta$ with {columns} $\{\btheta_1, \dots, \btheta_M\}$.
 
\textbf{Initialization:} Set initial weights $\bw_{k,0} = \cb{0}$  for all $k= 1, ..., N$.
\newline\textbf{Algorithm:}\hspace{2ex} For {each} instant $n \geq 1$, and for each agent $k$, update $\bw_{k,n-1}$:  
\begin{align}
	\bpsi_{k,n} 	&= (\bI_L-\mu\eta_2 \bP_{\bThetac})\bw_{k,n-1} 
				+  \mu\bx_{k,n}^*\big[d_k(n) -\bx_{k,n}\bw_{k,n-1}\big]
	\label{eq:Alg2.a} \\
	\bw_{k,n}   	&= \sum_{\ell\in\N{k}} a_{\ell k} \bP_{\bTheta}  \bpsi_{\ell,n} + \bP_{\bThetac} \bpsi_{k,n} 
	\label{eq:Alg2.b}
\end{align}
\caption{ATC diffusion LMS with node-specific subspace constraints (norm-bounded projections)} \label{algo:LMSmult2}
\end{algorithm}

%

\section{{Performance and convergence} analyses}
\label{sec:analyses}

In this section, we examine the convergence properties and network performance of the proposed adaptive strategies. We shall {first} {describe} a convergence framework for a family of distributed algorithms, where  Algorithms \ref{algo:LMSmult1} and~\ref{algo:LMSmult2} are {special cases}. {Quantities} specifically related to Algorithms \ref{algo:LMSmult1} or \ref{algo:LMSmult2} will be distinguished by superscripts $^{(1)}$ and $^{(2)}$, respectively.

In order to perform the analysis,  we collect information from across the network into block vectors and matrices. Let us denote by $\bw_n$ and $\bw^o$ the block weight vector at instant $n$ and the block optimum weight vector, both of size $LN\times 1$, that is
\begin{align}
             \bw_n &= \col\{\bw_{1,n}, \dots, \bw_{N,n}\}  \\
             \bw^o &= \col\{\bw_1^o, \dots, \bw_N^o\}
\end{align}
{We denote the difference between the optimum $\bw_k^o$ and the instantaneous estimate $\bw_{k,n}$ by:}
\begin{equation}
	\label{eq:v.tv}
	\bv_{k,n} = \bw^o_{k} - \bw_{k,n}
\end{equation}
We collect the weight error vectors $\bv_{k,n}$ from across all nodes into the block weight error vector:
\begin{equation}
        \bv_n = \col\{\bv_{1,n}, \dots, \bv_{N,n}\}
\end{equation}

 \assumption (Independent inputs) The regression vectors $\bx_{k,n}$ arise from a stationary random process that is temporally stationary, white, and independent over space with $\bR_{x,k} = \E\{\bx_{k,n}^*\bx_{k,n}^{\phantom{*}}\}>0$. A direct consequence {of this condition} is that $\bx_{k,n}$ is independent of $\bv_{\ell,m}$ for all $\ell$ and $m\leq n$.

\vspace{-5mm}
 \subsection{Mean weight behavior analysis}

The estimation error in~\eqref{eq:Alg1_1} and~\eqref{eq:Alg2.a} can be rewritten as a function of $\bv_{k,n}$:
\begin{equation}
	\label{eq:evx}
	d_{k}(n)-\bx_{k,n}\bw_{k,n-1} = z_{k}(n) + \bx_{k,n}\bv_{k,n-1}
\end{equation}
In what follows, {we first show} that the weight error update {relations} for both {Algorithms} \ref{algo:LMSmult1} and~\ref{algo:LMSmult2} {are} of the form:
\begin{equation}
            \label{eq:v.abbr0}
            \bv_{n} = \bB_{n}\bv_{n-1} - \mu\,\bg_{n} - \br,
\end{equation}
with $\bB_{n}$ an $LN\times LN$ time-dependent matrix,  $\bg_{n}$ an $LN\times 1$ zero-mean time-dependent vector, and $\br$ a constant $LN \times 1$ vector. Consequently, {it will be possible to represent their} mean weight behavior in the form of a state-transition equation with a bounded driving term:
\begin{equation}
           \label{eq:EvCom}
           \E\{ \bv_{n} \}= \bB\, \E\{ \bv_{n-1} \} - \br
\end{equation}
with $\bB = \E\{\bB_{n-1}\}$. Let $\bH_{x,n}$ be the block diagonal matrix of size $LN\times LN$, and $\bp_{zx,n}$ the vector of length $LN\times 1$, defined as follows:
 \begin{align}
	\bH_{x,n} 	&\triangleq \text{diag}\{\bx_{1,n}^*\bx_{1,n}^{\phantom{*}},\,\dots,\, \bx_{N,n}^*\bx_{N,n}^{\phantom{*}}\} \\
	\bp_{zx,n} &\triangleq \col\{z_{1}(n)\bx_{1,n}^*,\,\dots, \, z_{N}(n)\bx_{N,n}^*\}
 \end{align}
The expectation of $\bH_{x,n}$ and $\bp_{xz,n}$ are given by:
\begin{align}
              \bH_x  &\triangleq  \E\{\bH_{x,n}\} = \text{diag}\left\{\bR_{x,1},\,\dots,\bR_{N,2}\right\} \\
              \bp_{zx} &\triangleq \E\{\bp_{zx,n}\} = \cb{0}
\end{align}

\subsubsection{Mean weight behavior of Algorithm \ref{algo:LMSmult1}}

Define the intermediate weight error vector $\widetilde\bpsi_{k,n}$:
\begin{equation}
	\widetilde\bpsi_{k,n} = \bw^o_{k} - \bpsi_{k,n}
\end{equation}
{and collect these vectors from across all nodes into the block weight error vector:}
\begin{equation}
        \widetilde\bpsi_n = \col\{\widetilde\bpsi_{1,n}, \dots, \widetilde\bpsi_{N,n}\}
\end{equation}
Subtracting $\bw^o_k$ from both sides of the update relation~\eqref{eq:Alg1_1}, and using relation~\eqref{eq:evx}, leads to the update equation for $\widetilde\bpsi_{n}$:
\begin{equation}
	\label{eq:v1_2}
	\widetilde\bpsi_{n} = 
	(\bI_{LN} - \mu\bD_{\bS_{\bTheta}}\bH_{x,n}) \bv_{n-1}  - \mu\bD_{\bS_{\bTheta}}\bp_{zx,n}
\end{equation}
where $ \bD_{\bS_{\bTheta}} = \text{diag}\{\bS_{\bTheta}, \dots, \bS_{\bTheta}\}$ is an $LN\times LN$ block diagonal matrix with $\bS_{\bTheta}$ as diagonal entries. Let  ${\pA} = \bA \otimes \bI_L$. Defining $\bD_{\bP_{\bTheta}}$ and $ \bD_{\bP_{\bThetac}}$ as the $LN\times LN$ block diagonal matrices with $\bP_{\bTheta}$ and $\bP_{\bThetac}$ as diagonal entries, respectively,  equation~\eqref{eq:Alg1_2} can be written in vector form as:
\begin{equation}
	\bw_{n} = \big(\pA^\top \bD_{\bP_{\bTheta}} + \bD_{\bP_{\bThetac}}\big)\bpsi_{n}
\end{equation}
Subtracting $\bw^o$ from both sides of the above expression, we have:
\begin{equation}
           \bv_{n} = \big(\pA^\top \bD_{\bP_{\bTheta}} + \bD_{\bP_{\bThetac}}\big)\widetilde\bpsi_{n} 
           -\big(\pA^\top \bD_{\bP_{\bTheta}} + \bD_{\bP_{\bThetac}} - \bI_{LN}\big)\bw^o
\end{equation}
Combining this equation with~\eqref{eq:v1_2}, the weight error update relation can be written in a single expression:
\begin{align}
	\label{eq:Alg1.v}
		\bv_{n}  =\big(\!\pA^\top \bD_{\bP_{\bTheta}} + \bD_{\bP_{\bThetac}}\big)  \big[(\bI_{LN} 
		- \mu\bD_{\bS_{\bTheta}}\bH_{x,n} ) \bv_{n-1}
		- \mu\bD_{\bS_{\bTheta}}\bp_{zx,n}\big]- (\pA^\top - \bI_{LN}) \bD_{\bP_{\bTheta}}\bw^o
\end{align}
Now we denote several terms in the weight error expression~\eqref{eq:Alg1.v} by:
 \begin{align}
	\bB_n^{(1)} &=  \big(\pA^\top \bD_{\bP_{\bTheta}} + \bD_{\bP_{\bThetac}}\big)  (\bI_{LN} 
	- \mu\bD_{\bS_{\bTheta}}\bH_{x,n} )    \\
	\bg_n^{(1)} & =  \big(\pA^\top \bD_{\bP_{\bTheta}} + \bD_{\bP_{\bThetac}}\big)\bD_{\bS_{\bTheta}}\bp_{zx,n} \\
	\br^{(1)} &= (\pA^\top - \bI_{LN}) \bD_{\bP_{\bTheta}}\bw^o,
 \end{align}
 and the associated expected values:
 \begin{align}
	\bB^{(1)} &\triangleq \E\{\bB_n^{(1)}\} \nonumber \\&= \big(\pA^\top \bD_{\bP_{\bTheta}} 
	+ \bD_{\bP_{\bThetac}}\big)  (\bI_{LN} - \mu\bD_{\bS_{\bTheta}}\bH_{x} )   \\
	\bg^{(1)} &\triangleq\E\{\bg_n^{(1)}\} = \bf{0}
 \end{align}
With the above notation, the weight error update relation~\eqref{eq:Alg1.v} can be written as:
\begin{equation}
            \label{eq:Alg1.v.abbr}
            \bv_{n} = \bB_n^{(1)} \, \bv_{n-1} - \mu\,\bg_n^{(1)} - \br^{(1)}
\end{equation}
Taking the  expectation on  both sides of~\eqref{eq:Alg1.v.abbr}, and using Assumption 1, we {arrive at} the mean weight behavior {for} Algorithm~\ref{algo:LMSmult1}:
\begin{equation}
	\label{eq:Ev.abbr.alg1}
	\E\{\bv_{n}\}  = \bB^{(1)}  \, \E\{\bv_{n-1}\} - \br^{(1)}
\end{equation}

\subsubsection{Mean weight behavior of Algorithm \ref{algo:LMSmult2}}

{
Subtracting $\bw^o_k$ from both sides of the update relation~\eqref{eq:Alg2.a}, and using relation~\eqref{eq:evx}, yields:
\begin{align}
	\label{eq:v1_2-alg2}
	\widetilde\bpsi_{n}
	&=(\bI-\mu\eta_2\bD_{\bP_{\bThetac}} - \mu\bH_{x,n}) \bv_{n-1} 
	- \mu(\bp_{zx,n} - \eta_2\bD_{\bP_{\bThetac}} \bw^o)
\end{align}
Subtracting $\bw^o$ from both sides of~\eqref{eq:Alg2.b}, we have:
\begin{equation}
           \bv_{n} =\! \big(\pA^\top \bD_{\bP_{\bTheta}} + \bD_{\bP_{\bThetac}}\big)\widetilde\bpsi_{n}
           -\big(\pA^\top \bD_{\bP_{\bTheta}} + \bD_{\bP_{\bThetac}} \!- \bI_{LN}\big)\bw^o
\end{equation}
Combining this equation with~\eqref{eq:v1_2-alg2}, the weight error update relation can be written in a single expression:
\begin{equation}
	\label{eq:v}
	\begin{split}
		\bv_{n}  =& \big(\pA^\top \bD_{\bP_{\bTheta}} \!+\! \bD_{\bP_{\bThetac}}\big)  \big[ (\bI_{LN} \! \
		\!-\!\mu\eta_2 \bD_{\bP_{\bThetac}} \!\!-\!\mu\bH_{x,n} ) \bv_{n-1}  \\
		&- \mu(\bp_{zx,n}- \eta_2\bD_{\bP_{\bThetac}}\bw^o)\big] 
		 - (\pA^\top - \bI_{LN}) \bD_{\bP_{\bTheta}}\,\bw^o
	\end{split}
\end{equation}
where we used the fact that $\bI_{LN}=\bD_{\bP_{\bTheta}} + \bD_{\bP_{\bThetac}}$. Next, we denote several terms in the weight error expression~\eqref{eq:v} by:
\begin{align}
	\!\!\!\! \bB_n^{(2)} &=  \big(\pA^\top \bD_{\bP_{\bTheta}} \!\!+\!\! \bD_{\bP_{\bThetac}}\!\big)  (\bI_{LN} \!-\!\mu\eta_2\bD_{\bP_{\bThetac}} \!\!-\! \mu\bH_{x,n}) \\
	\!\!\!\!\bg_n^{(2)} & =  \big(\pA^\top \bD_{\bP_{\bTheta}} + \bD_{\bP_{\bThetac}}\big)\bp_{zx,n} \\
	\!\!\!\!\br^{(2)} 	  &=  (\pA^\top - \bI_{LN}\!)\bD_{\bP_{\bTheta}}\bw^o 
	 - \mu \eta_2\big(\!\pA^\top\bD_{\bP_{\bTheta}}
	+\bD_{\bP_{\bThetac}}\big){\bD_{\bP_{\bThetac}}\bw^o} \!
\end{align}
and the associated expected values:
 \begin{align}
	\!\!\!\! \bB^{(2)} &\triangleq \E\{\bB_n^{(2)}\} \nonumber \\ 
	\!\!\!\! & = \big(\pA^\top \bD_{\bP_{\bTheta}} 
	\!+\! \bD_{\bP_{\bThetac}}\big)  (\bI_{LN} \!-\!\mu\eta_2\bD_{\bP_{\bThetac}} \!-\! \mu\bH_{x})   \\
	\!\!\!\! \bg^{(2)} &\triangleq\E\{\bg_n^{(2)}\} = \bf{0}
 \end{align}
With the above notation, the weight error update relation~\eqref{eq:v} can be written as:
\begin{equation}
            \label{eq:v.abbr}
            \bv_{n} = \bB_n^{(2)} \, \bv_{n-1} - \mu\,\bg_n^{(2)} - \br^{(2)}
\end{equation}
Taking the  expectation on  both sides of~\eqref{eq:v.abbr}, and using Assumption 1, we get the mean weight behavior of Algorithm~\ref{algo:LMSmult2}:
\begin{equation}
	\label{eq:Ev.abbr.alg2}
	\E\{\bv_{n}\}  = \bB^{(2)}  \, \E\{\bv_{n-1}\} - \br^{(2)}
\end{equation}}

\subsubsection{Stability in the mean}

{The mean-weight error recursions}~\eqref{eq:Ev.abbr.alg1} and~\eqref{eq:Ev.abbr.alg2} are of the same form as~\eqref{eq:EvCom}. The convergence of such recursive state-transition equations, with a bounded driving term, is determined by the stability of matrix $\bB$. Algorithm parameters should be chosen to satisfy the mean stability condition $\rho(\bB)<1$, where $\rho(\cdot)$ denotes spectral radius of its matrix argument. In this case, the bias of the algorithm {will be given by:}
\begin{equation}
	\label{eq:Ev_inf}
	\lim_{n\rightarrow\infty} \E\{\bv_n\} = -(\bI_{LN}-\bB)^{-1}\br
\end{equation}
We shall now establish two results that provide ranges for selecting the step size $\mu$ {to ensure} convergence in the mean {for} each algorithm.

\theorem (Stability in the mean {for} Algorithm \ref{algo:LMSmult1}) Assume data model~\eqref{eq:datamodel} and Assumption 1 hold. We select a doubly stochastic matrix $\bA$. Assume $\{\bTheta,\bThetac\}$ forms an orthonormal basis of $\R^L$. Then, for any initial condition, Algorithm \ref{algo:LMSmult1} asymptotically converges in the mean if the  step-size satisfies:
\begin{equation}
	\label{eq:stp1}
	{0 <  \mu < \frac{2}{\max_k \lambda_{\max}(\bR_{x,k})}}
\end{equation}
where $\lambda_{\max}(\cdot)$ denotes the maximum eigenvalue of its matrix argument.
\begin{proof}
The convergence of~\eqref{eq:Ev.abbr.alg1} is determined by the stability of matrix $\bB^{(1)}$. The required mean stability condition is met by selecting $\mu$ so that:
\begin{equation}
	\rho\big((\pA^\top \bD_{\bP_{\bTheta}} + \bD_{\bP_{\bThetac}})  (\bI_{LN} - \mu\bS_\Theta\bH_{x})\big) < 1
\end{equation}
Let $\bx=\col\{\bx_{1}, \dots, \bx_{N}\}$ be any block vector of size $LN \times 1$. We have:
\begin{align}
	 \big\|\big(\pA^\top \bD_{\bP_{\bTheta}} + \bD_{\bP_{\bThetac}}\big)  \bx\big\|^2 \label{eq:proof1}  
	=& \sum_{i=1}^N \Big \|\sum_{j=1} ^N a_{ji}  \, \bP_{\bTheta}\bx_j + \bP_{\bThetac} \bx_i \Big\|^2 \\
	=& \sum_{i=1}^N \Big(\Big \| \sum_{j=1} ^N a_{ji}  \, \bP_{\bTheta}\bx_j  \Big \|^2 + \big \|\bP_{\bThetac} \bx_i \big\|^2 \Big)  \label{eq:proof2}
\end{align}
Given that $\bA$ is left stochastic, namely, $\sum_{j=1}^N a_{ji}= 1$ with $a_{ji} \geq 0$, Jensen's inequality guarantees:
\begin{equation}
	\Big \|\sum_{j=1} ^N a_{ji}  \, \bP_{\bTheta}\bx_j  \Big \|^2 \leq \sum_{j=1}^N a_{ji} \big \| \bP_{\bTheta}\bx_j \big \|^2
\end{equation}
Consequently, the quantity in \eqref{eq:proof2} can be upper-bounded as follows:
\begin{align}
	\sum_{i=1}^N \Big(  \Big \|  \sum_{j=1} ^N a_{ji}  \, \bP_{\bTheta} \bx_j  \Big \|^2 + \big \|\bP_{\bThetac} \bx_i \big\|^2 \Big)  
	&\leq   \sum_{i=1}^N \sum_{j=1}^N a_{ji} \big \| \bP_{\bTheta}\bx_j \big \|^2  + \sum_{i=1}^N \big \|\bP_{\bThetac} \bx_i \big\|^2  \\ 
	& \stackrel{(a)}{=} \sum_{j=1}^N  \big \| \bP_{\bTheta} \bx_j \big \|^2  + \sum_{i=1}^N \big \|\bP_{\bThetac} \bx_i \big\|^2 \\ 
	& = \|\bx\|^2
\end{align}
where for step (a) we use that $\bA$ is right stochastic, namely, $\sum_{i=1}^N a_{ji}= 1$. We conclude that:
\begin{equation}
	\label{eq:condproj}
	\big\|\pA^\top \bD_{\bP_{\bTheta}} + \bD_{\bP_{\bThetac}}\big\| \leq 1
\end{equation}
We know that the spectral radius of any matrix $\bX$ satisfies $\rho(\bX)\leq\|\bX\|$, for any induced norm. Then we have:
\begin{align}
	\rho\big((\pA^\top \bD_{\bP_{\bTheta}} + \bD_{\bP_{\bThetac}})  (\bI_{LN} - \mu\bS_\Theta\bH_{x})\big)
	&\leq\big\|\pA^\top \bD_{\bP_{\bTheta}} + \bD_{\bP_{\bThetac}}\big\|\,\big\|\bI_{LN} - \mu\bS_\Theta\bH_{x}\big\| \\
	&\stackrel{{\eqref{eq:condproj}}}{\leq}\big\|\bI_{LN} - \mu\bS_\Theta\bH_{x}\big\|
\end{align}
The mean stability condition is thus met by selecting $\mu$ so that: $\big\|\bI_{LN} - \mu\bS_\Theta\bH_{x}\big\|<1$. In the case where
$\{\bTheta,\bThetac\}$ forms an orthonormal basis of $\R^L$, then $\bS_\Theta=\bI_{L}$. This leads us to the condition in~\eqref{eq:stp1}.

\end{proof}

\theorem (Stability in the mean for Algorithm \ref{algo:LMSmult2}) Assume data model~\eqref{eq:datamodel} and Assumption 1 hold. We select a doubly stochastic matrix $\bA$. Then, for any initial condition, Algorithm 2 asymptotically converges in the mean if the step-size satisfies:
\begin{equation}
	\label{eq:stp1.alg2}
	 0 <  \mu < \frac{2}{\max_k \lambda_{\max}( \eta_2 \bP_{\bThetac} + \bR_{x,k})}
\end{equation}

\begin{proof}
The convergence of~\eqref{eq:Ev.abbr.alg2} is determined by the stability of matrix $\bB^{(2)}$. Considering that:
\begin{equation}
      \begin{split}
	\rho\big((\pA^\top \bD_{\bP_{\bTheta}} + \bD_{\bP_{\bThetac}})(\bI_{LN} -\mu\eta_2  \,\bD_{\bP_{\bThetac}} - \mu\bH_{x} )\big)
	\leq
	\big\|\bI_{LN} -\mu\eta_2  \,\bD_{\bP_{\bThetac}} - \mu\bH_{x}\big\|
	\end{split}
\end{equation}
since $\|\pA^\top \bD_{\bP_{\bTheta}} + \bD_{\bP_{\bThetac}}\| \leq 1$, the mean stability condition is met by selecting $\mu$ so that $\big\|\bI_{LN} -\mu\eta_2  \,\bD_{\bP_{\bThetac}} - \mu\, \bH_{x} \big\|<1$. This leads us to the condition in \eqref{eq:stp1.alg2}. Furthermore, by Weyl's theorem, we have $\lambda_{\max}( \eta_2 \bP_{\bThetac} + \bR_{x,k})\leq\eta_2 + \lambda_{\max}(\bR_{x,k})$ since $\bP_{\bThetac}$ and $\bR_{x,k}$ are Hermitian matrices and $\lambda_{\max}(\bP_{\bThetac})=1$. This leads to the sufficient condition:
\begin{equation}
	  0 <  \mu < \frac{2}{\eta_2+\max_k \lambda_{\max}(\bR_{x,k})}
\end{equation}
\end{proof}

\vspace{-6mm}
\subsection{Mean-square error behavior analysis}
 
We now study the mean-square error behavior of  Algorithms~\ref{algo:LMSmult1} and~\ref{algo:LMSmult2}. To this end, we consider the general update relation~\eqref{eq:EvCom} since both algorithms are of this form. 
 {From~\eqref{eq:v.abbr0}, the  squared norm $\|\bv_n\|^2_{\bSig}$ of} the weight vector $\bv_{n}$ weighted by any {positive} semi-definite matrix $\bSig$, {i.e., $\|\bv_n\|^2_{\bSig} = \bv_n^* \bSig \bv_n$}, satisfies the following relation:
 {
\begin{equation}
	\label{eq:bvn}
        \begin{split}
        \|\bv_n\|^2_{\bSig}  &=  \|\bv_{n-1}\|^2_{\bB_n^*\bSig\bB_n} - \mu^2 \|\bg_n\|_{\bSig} + \|\br\|^2_{\bSig} 
                                        - 2 \,\text{Re}\{ \br^*\bSig\bB_n\bv_{n-1} \} - 2 \mu \,\text{Re}\{\bg_n^*\bSig(\bB_n\bv_{n-1}-\br)\}
        \end{split}
\end{equation}
Under the independence assumption, and considering that $\bg_n$ includes the zero-mean noise term $z_n$ which is independent of any other signal, taking expectations of both sides of \eqref{eq:bvn} leads to:}
 \begin{equation}
	\label{eq:EvSig}
	\begin{split}
          \E\{\|\bv_{n}\|^2_{\bSig}\} &= \E\{\|\bv_{n-1}\|^2_{\bSig'}\} 
	+ \mu^2\, \tr\!\left\{ \bSig  \E\{\bg_n\bg_n^*\} \right\} 
	+  \|\br\|^2_{\bSig}
	-2 \,\text{Re}\big\{\E\{ \br^*\bSig\bB_n\bv_{n-1} \}\big\}
	\end{split}
 \end{equation}
In the above expression, $\bSig$ is any positive semi-definite matrix that the user is free to choose in order to derive different performance metrics, and $\bSig' = \E\{ \bB_n^*\bSig\bB_n^{\phantom{*}}\}$. Let $\bG$ be the expected value of $\E\{\bg_n\bg_n^*\}$ in the second term on the RHS of~\eqref{eq:EvSig}. For the two presented algorithms, $\bG$ is respectively given by:
\begin{align}
           \bG^{(1)} &=(\pA^\top \bD_{\bP_{\bTheta}} + \bD_{\bP_{\bThetac}})
           \bD_{\bS_{\bTheta}} \text{diag}\{\sigma_{z,1}^2 \bR_{x,1},  \dots,\sigma_{z,N}^2 \bR_{x,N} \} \,
           \bD_{\bS_{\bTheta}}^* (\pA^\top \bD_{\bP_{\bTheta}} \!+\! \bD_{\bP_{\bThetac}})^* \\
           \bG^{(2)} &= \text{diag}\{\sigma_{z,1}^2 \bR_{x,1},\dots,\sigma_{z,N}^2 \bR_{x,N}\}
\end{align}
With $\bG$, equation~\eqref{eq:EvSig} is expressed as:
 \begin{equation}
         \begin{split}
          \label{eq:EvSig2}
          \E\{\|\bv_{n}\|^2_{\bSig}\} &= \E\{\|\bv_{n-1}\|^2_{\bSig'}\} + \mu^2 \tr \left\{ \bSig \bG \right\} +  \|\br\|^2_{\bSig} \\
                                                     & -2 \,\text{Re}\big\{ \br^*\bSig\,  \bB\, \E\{\bv_{n-1} \}  \big\}
          \end{split}
 \end{equation}
Vectorizing matrices $\bSig$ and $\bSig'$ by $\bsig=\vc(\bSig)$ and $\bsig' = \vc(\bSig')$, it can be verified that:\begin{equation}
         \bsig' = \bK\bsig
\end{equation}
where the $(LN)^2 \times (LN)^2$ matrix $\bK$ is given by:
\begin{equation}
         \label{eq:Kapp}
         \begin{split}
        \bK = \; \E\{\bB_n^\top \otimes \bB_n^*\}  \approx  \; \bB^\top \otimes \bB^*
        \end{split}
\end{equation}
The above approximation can be used provided that the step size is sufficiently small so that the influence of the second-degree term in $\mu$ can be  neglected~\cite{Sayed2013intr}. Equation~\eqref{eq:EvSig2} can then be expressed as:
\begin{equation}
	\label{eq:EvSig3}
	\begin{split}
		\E\{\|\bv_{n}\|^2_{\bsig}\} &= \E\{\|\bv_{n-1}\|^2_{\bK\bsig}\} + \bs_{n-1}^\top\,\bsig
	\end{split}
 \end{equation}
 where we use the notation $\|\!\cdot\! \|_\Sigma$ and $\|\!\cdot\! \|_\sigma$ interchangeably, and
 \begin{equation}
 	\label{eq:def-s}
	\bs_{n-1}= \vc\Big(\mu^2\,\bG + \br \br^*- 2 \, \text{Re}\big\{\bB\,\E\{ \bv_{n-1}\} \br^* \big\}\Big)
 \end{equation}

\theorem (Mean-square stability) Assume data model~\eqref{eq:datamodel} and Assumption 1 hold. Assume further that the step size~$\mu$ is sufficiently small to guarantee the stability in the mean of the algorithms, and to ensure that \eqref{eq:EvSig3} can be used as a reasonable representation for the evolution of the (weighted) mean-square error.  Mean-square stability of cooperative algorithms characterized by~\eqref{eq:v.abbr0} requires the step-size $\mu$ to be chosen such that it ensures the stability of matrix $\bK$ (in addition to the mean stability condition  $\rho(\bB)<1$).

\begin{proof}
Iterating~\eqref{eq:EvSig3} starting from $n=0$, we find that
\begin{equation}
         \label{eq:EVn0}
                 \E\{\|\bv_{n}\|^2_{\bsig}\} = \|\bv_0\|^2_{\bK^{n}\bsig} +  \sum_{i=1}^n  \bs_{n-i}^\top  \bK^{i-1} \bsig
\end{equation}
with initial condition $\bv_0=\bw^o-\bw_0$. Provided that $\bK$ is stable, the first term on the RHS of~\eqref{eq:EVn0} converges to zero as $n\rightarrow\infty$. We know from~\eqref{eq:EvCom} that $\E\{\bv_n\}$ is bounded because~\eqref{eq:EvCom} is a BIBO stable recursion with a bounded driving term~$\br$. The second term on the RHS of~\eqref{eq:EVn0} then converges as $n\rightarrow\infty$. We conclude that $\E\{\|\bv_{n}\|^2_{\bsig}\}$ converges to a bounded value as $n\rightarrow\infty$, and the algorithm is mean-square stable.
 \end{proof}

\theorem (Transient MSD) Consider a sufficiently small step size $\mu$ to ensure mean and mean-square stabilities. The MSD learning curve $\zeta_n \!=\! \frac{1}{N}\E\{\|\bv_n\|^2\}$ of the cooperative algorithms characterized by~\eqref{eq:v.abbr0}, obtained by setting $\bSig = \frac{1}{N}\bI_{LN}$, evolves according to the following recursion for~$n\geq 1$:
\begin{align}
         \zeta_{n}  &= \zeta_{n-1}+\frac{1}{N} \big[(\bgam_{n-1}+\bs_{n-1})^\top\vc(\bI_{LN}) -\|\bv_0\|^2_{(\bI_{(LN)^2}-\bK)\bK^{n-1}\bsig}\big]     \label{eq:TransMSD1} \\
               \bgam_{n} &= \bK^\top \bgam_{n-1} + (\bK-\bI_{(LN)^2})^\top \bs_{n-1} \label{eq:TransMSD2}
\end{align}
with initial conditions $\zeta_0 = \frac{1}{N}\|\bv_0\|^2$ and $\bgam_0  =\cb{0}$.
\begin{proof}

Comparing~\eqref{eq:EVn0} at instants $n$ and $n-1$, we can relate  $\E\{\|\bv_{n}\|^2_{\bsig}\}$  and  $\E\{\|\bv_{n-1}\|^2_{\bsig}\} $ as follows:
\begin{equation}
	\label{eq:EVrec_sig}
	\begin{split}
	\E\{\|\bv_{n}\|^2_{\bsig}\} &= \E\{\|\bv_{n-1}\|^2_{\bsig}\}   
	- \|\bv_0\|^2_{(\bI_{(LN)^2}-\bK)\bK^{n-1}\bsig}  +\sum_{i=1}^n\bs_{n-i}^\top\bK^{i-1}\bsig 
	- \sum_{i=1}^{n-1}\bs_{n-1-i}^\top\bK^{i-1}\bsig\\
	&= \E\{\|\bv_{n-1}\|^2_{\bsig}\}   - \|\bv_0\|^2_{(\bI_{(LN)^2}-\bK)\bK^{n-1}\bsig} 
	\!+\!\bs_{n-1}^\top\bsig \!+\! \sum_{i=2}^n\bs_{n-i}^\top\bK^{i-1}\bsig \!-\! \sum_{i=1}^{n-1}\bs_{n-1-i}^\top\bK^{i-1}\bsig
	\end{split}
\end{equation}
Introducing the notation
\begin{equation}
	\bgam_{n-1} = \Big[\sum_{i=2}^n\bs_{n-i}^\top\bK^{i-1}  - \sum_{i=1}^{n-1}\bs_{n-1-i}^\top\bK^{i-1}\Big]^\top
\end{equation}
we can reformulate the recursive expression~\eqref{eq:EVrec_sig} as follows:
\begin{align}
	\E\{\|\bv_{n}\|^2_{\bsig}\} &= \E\{\|\bv_{n-1}\|^2_{\bsig}\}  - \|\bv_0\|^2_{(\bI_{(LN)^2}-\bK)\bK^{n-1}\bsig}  + (\bgam_{n-1} +\bs_{n-1})^\top\!\bsig\\
	\bgam_{n} &= \bK^\top \bgam_{n-1} +    (\bK-\bI_{(LN)^2})^\top \bs_{n-1}
\end{align}
with $\bgam_0 = \cb{0}$. To derive the transient curve for the MSD, replace $\bsig$ by $\frac{1}{N}\vc\{\bI_{LN}\}$.
\end{proof}

\corollary (Steady-state MSD)  If the step size is chosen sufficiently small to ensure mean and mean-square convergence, then the steady-state MSD, defined as $\zeta_\infty = \lim_{n\rightarrow\infty}\zeta_n$, is given by:
\begin{equation}
        \label{eq:MSD}
        \zeta_\infty=  \frac{1}{N}\, \bs_\infty^\top\,   (\bI_{(LN)^2} - \bK)^{-1} \vc(\bI_{LN})
\end{equation}
with $\bs_\infty=\lim_{n\rightarrow\infty}\bs_n$ determined by \eqref{eq:def-s}, using $\E\{\bv_\infty\}=\lim_{n\rightarrow\infty}\E\{\bv_n\}$ determined by~\eqref{eq:Ev_inf}.
\begin{proof} 

From expression~\eqref{eq:EvSig3}, we get:
\begin{equation}
	\label{eq:stablepoint}
        \lim_{n\rightarrow\infty}\E\{\|\bv_n\|^2_{(\bI_{(LN)^2}-\bK)\bsig}\} =   \bs^\top_\infty\bsig
\end{equation}
Observe that the MSD calculation requires us to choose $\bsig$ that satisfies:
\begin{equation}
	\label{eq:sig_MSD}
	(\bI_{(LN)^2} - \bK) \,\bsig = \frac{1}{N} \vc(\bI_{LN})
\end{equation}
This leads to expression~\eqref{eq:MSD}.
\end{proof}

\vspace{-2mm}
\section{Simulations}
In this section, we report simulation results that {illustrate the} theoretical results. All agents were initialized with zero parameter vectors $\bw_{k,0}=\cb{0}$ for all $k$. Simulated curves were obtained by averaging  over 100 runs as we obtained sufficiently smooth curves to check the consistency with theoretical results.

\vspace{-4mm}
\subsection{Algorithm validation}

\begin{figure*}[!t]
	\subfigure[Network topology.]{
 	\label{fig:Top}
   	\begin{minipage}[c]{.5\linewidth}
   		\centering
      		\includegraphics[trim = 0mm -17mm 0mm -17mm, clip, scale=0.4]{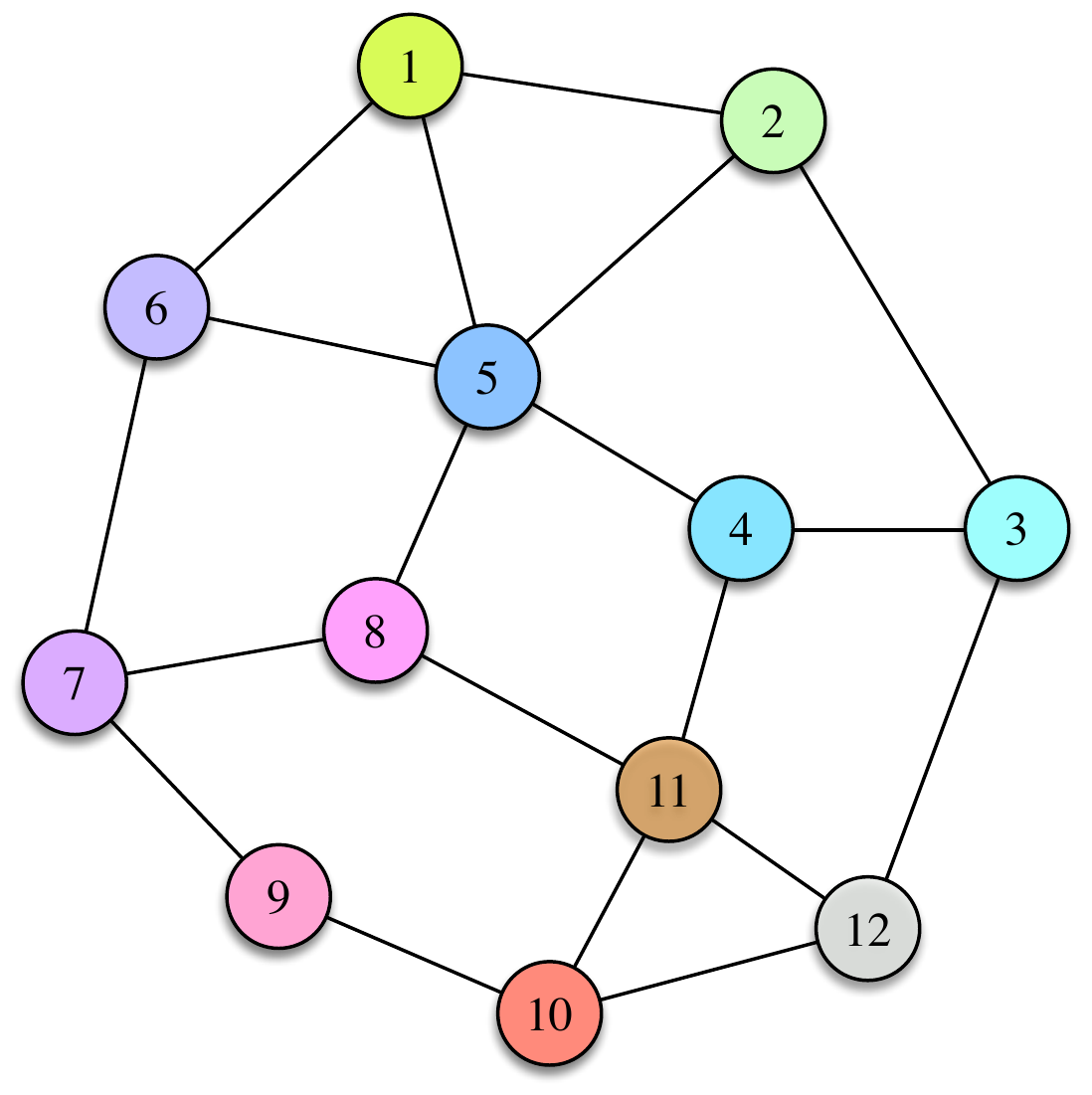}
   	\end{minipage}} \hfill
	\subfigure[Agent input and noise variances.]{
 	\label{fig:Var}
   		\begin{minipage}[c]{.5\linewidth}
   		\centering
      		\includegraphics[trim = 0mm 10mm -20mm 0mm, clip, scale=0.35]{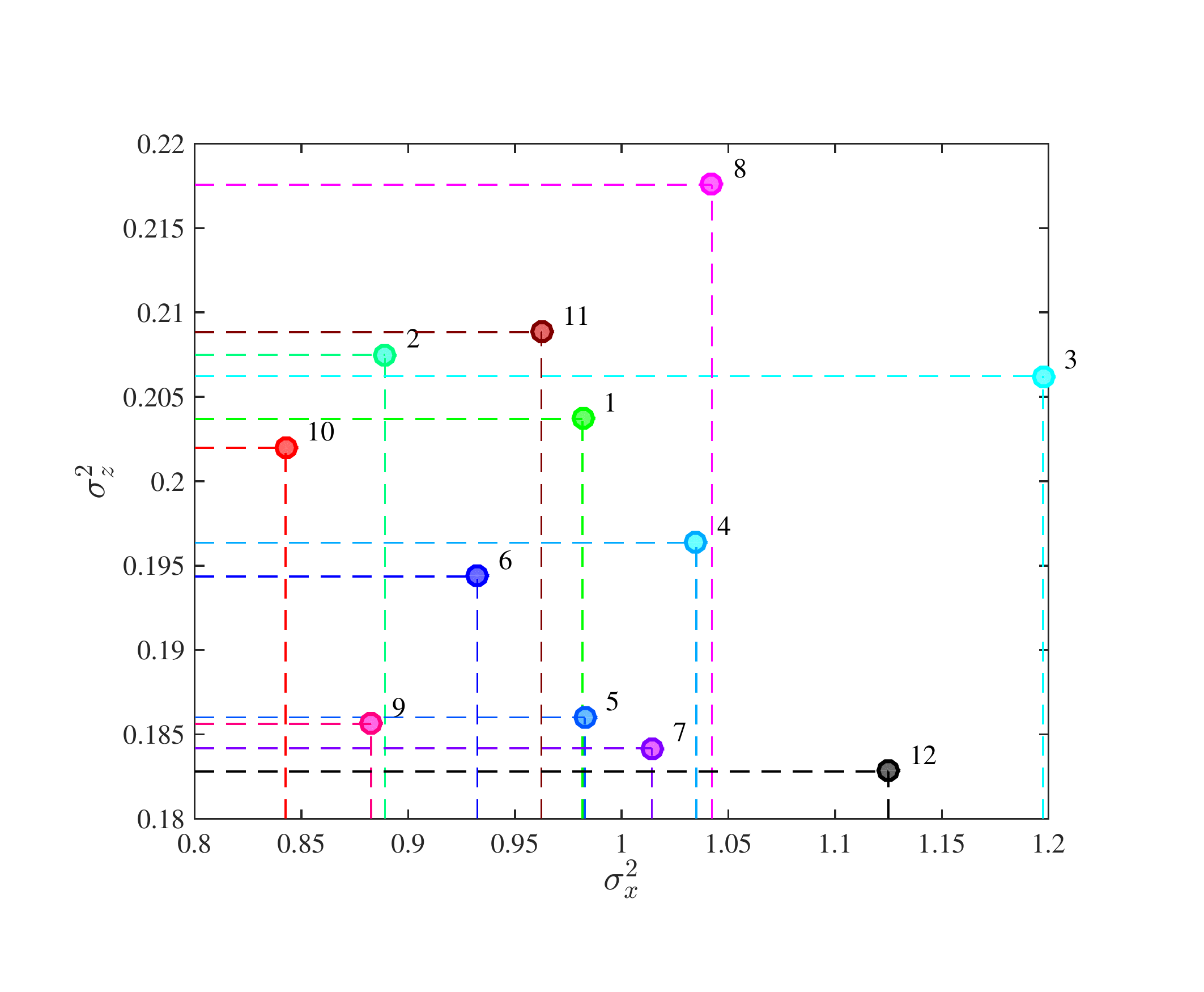}
   	\end{minipage}} 
	\vspace{-5mm}
	\caption{Network topology and input-noise variances.}
	\label{fig:Network}
	\vspace{-5mm}
\end{figure*}

We considered a network consisting of 12 agents with interconnections shown in Fig.~\ref{fig:Top}.  The parameter vectors to be estimated were of length $L = 5$. The input data $\bx_{k,n}$ were generated from circularly-symmetric zero-mean complex Gaussian distributions. White input data were considered first, by setting:
\begin{equation}
	\bR_{x,k} = \sigma_{x,k}^2\, \bI_5
\end{equation}
Next, correlated input data, characterized by the following covariance matrix, were considered:
\begin{equation}
       \begin{split}
	\bR_{x,k} = \sigma_{x,k}^2 \times 
	\left(\begin{array}{ccccc}
		1 & -.4+.3j & .2-.1j & .1-.05j & .02+.02j \\
		-.4-.3j & 1 & -.4+.3j & .2-.1j & .1-.05j \\
		.2+.1j & -.4-.3j & 1 & -.4+.3j & .2-.1j \\
		.1+.05j & .2+.1j & -.4-.3j & 1 & -.4+.3j \\
		.02-.02j & .1+.05j & .2+.1j & -.4-.3j & 1
\end{array}\right)
        \end{split}
\end{equation}
with ${ j} = \sqrt{-1}$ the imaginary unit. The modeling noises $z_{k,n}$ were i.i.d. zero-mean circularly-symmetric Gaussian variables, independent of any other signals. The variances $\sigma^2_{x,k}$ and $\sigma^2_{z,k}$ were sampled from $\cp{U}(0.8, 1.2)$ and $\cp{U}(0.18,0.22)$, respectively. Their values are shown in Fig.~\ref{fig:Var}. We considered two sets of subspace basis vectors. The first set is the standard basis:
\begin{equation}
	\bTheta_1 =[\cb{e}_1, \cb{e}_2, \dots, \cb{e}_M], 
\end{equation}
where $\cb{e}_i$ denotes a vector of length $N$ with $1$ at the $i$th entry and $0$ otherwise.
Its {orthogonal} complementary subspace is spanned by $\bTheta_{1,\perp} = [\cb{e}_{M+1}, \dots, \cb{e}_{L} ]$. This setup can be interpreted as a variable selection process for information exchange, where the first $M$ entries of the optimal parameter vectors are identical across the network. Parameter $M$ was set to $3$. The second set of basis vectors is a complex Vandermonde matrix:
\begin{equation}
           \label{eq:arrayA}
           \bTheta_2 = \left( \begin{array}{cccc} 1 & 1& \cdots &1 \\ e^{-j\psi_1} & e^{-j\psi_2} & \cdots & e^{-j\psi_M} \\ \cdots & \cdots & \cdots& \cdots \\ e^{-j(L-1)\psi_1} & e^{-j(L-1)\psi_2} & \cdots & e^{-j(L-1)\psi_M}\end{array}\right)
\end{equation}
with $\psi_k = \frac{2\pi d}{\lambda_{\rm o}} \sin \theta_k$. Matrix $\bTheta_2$ can represent the array manifold of a uniform linear array (ULA) with inter-element space~$d$,  operating at wavelength $\lambda_{\rm o}$ with impinging signal directions {of angles} $\theta_k$.  Parameter $M$ was set to $3$, with $\theta_1= \frac{\pi}{6}$, $\theta_2= \frac{\pi}{4}$, $\theta_3= \frac{\pi}{3}$ and $d=\frac{\lambda_{\rm o}}{2}$. {We considered three {settings to validate the theoretical results.}}



\begin{figure*}[!t]
	\hspace{-0cm}
	\subfigure[Using $\bTheta_1$.]{
 	\label{fig:Alg1THT1}
   	\begin{minipage}[c]{.3\linewidth}
   		\centering
      		\includegraphics[trim = 0mm 10mm 0mm 0mm, clip, scale=0.35]{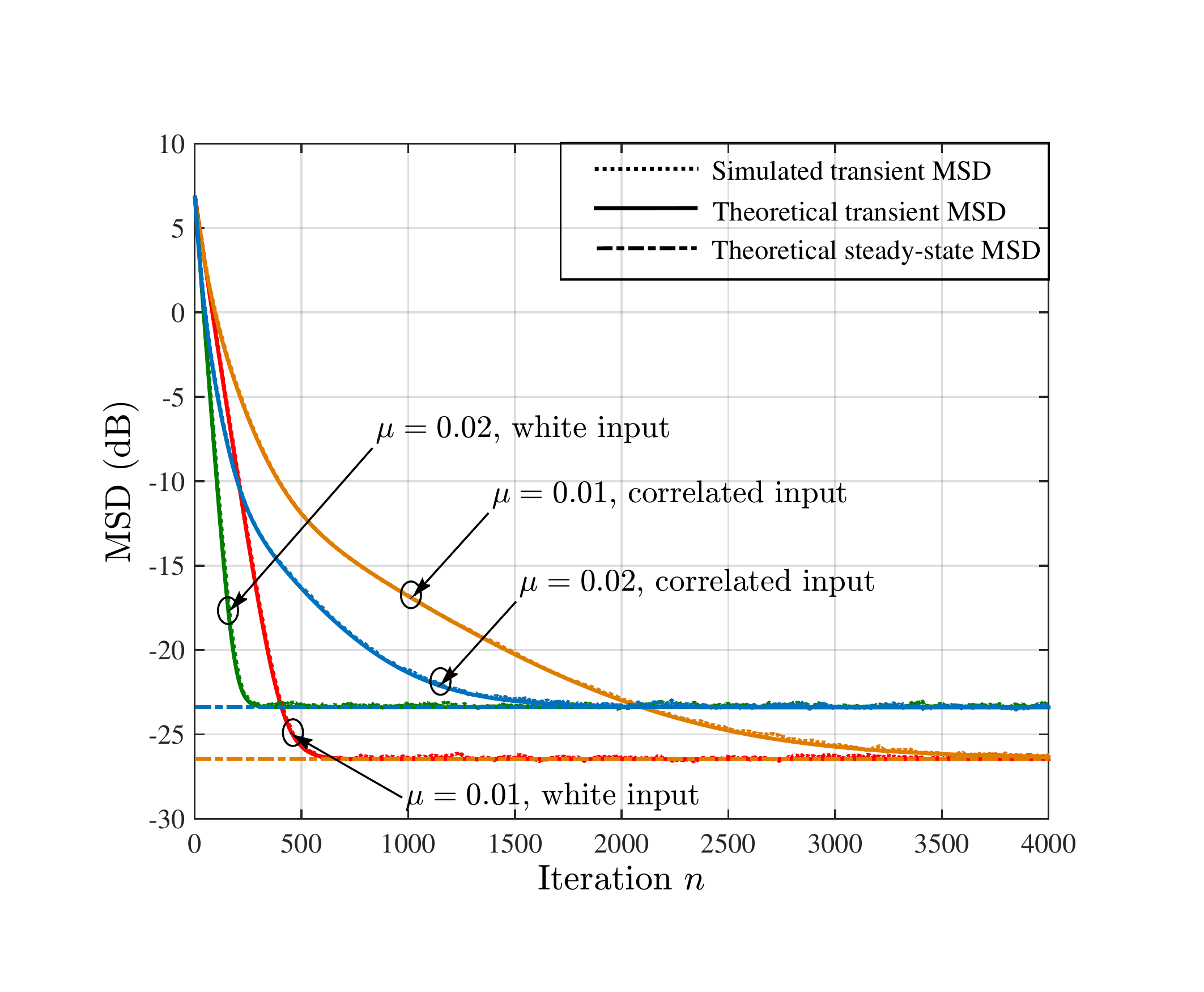}
   	\end{minipage}} \hspace{30mm}
	\subfigure[Using $\bTheta_2$.]{
 	\label{fig:Alg2THT1}
   		\begin{minipage}[c]{.3\linewidth}
      		\includegraphics[trim = 0mm 10mm 0mm 0mm, clip, scale=0.35]{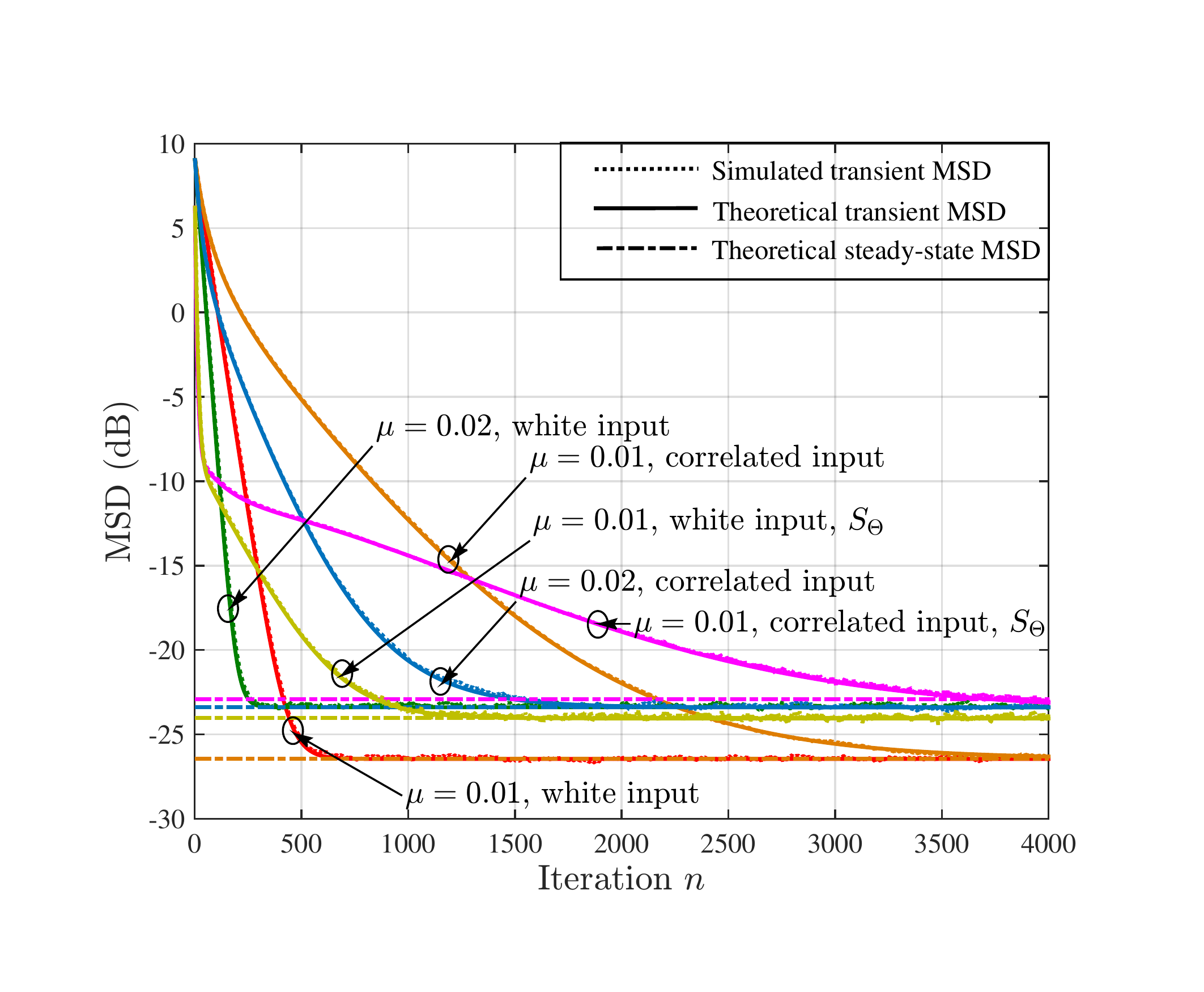}
   	\end{minipage}}  \hspace{2.5mm}
	\vspace{-2mm}
	\caption{Learning curves and model validation of  Algorithm 1 with different settings.}
	\label{fig:Alg1}
	\vspace{-8mm}
\end{figure*}

\begin{figure*}[!t]
	\hspace{-0cm}
	\subfigure[Using $\bTheta_1$.]{
 	\label{fig:Alg1THT1}
   	\begin{minipage}[c]{.3\linewidth}
   		\centering
      		\includegraphics[trim = 0mm 10mm 0mm 0mm, clip, scale=0.35]{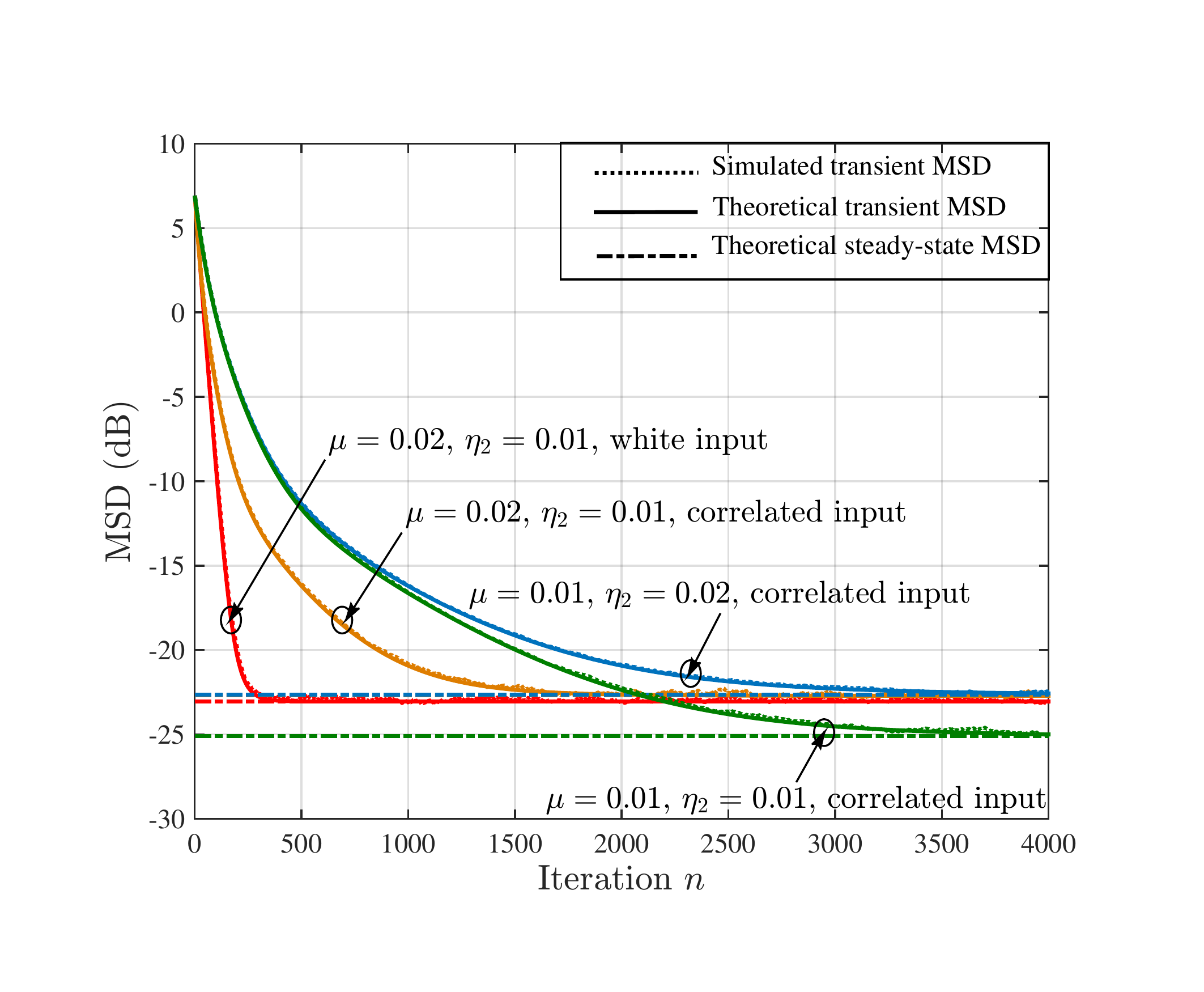}
   	\end{minipage}} \hspace{30mm}
	\subfigure[Using $\bTheta_2$.]{
 	\label{fig:Alg2THT1}
   		\begin{minipage}[c]{.3\linewidth}
      		\includegraphics[trim = 0mm 10mm 0mm 0mm, clip, scale=0.35]{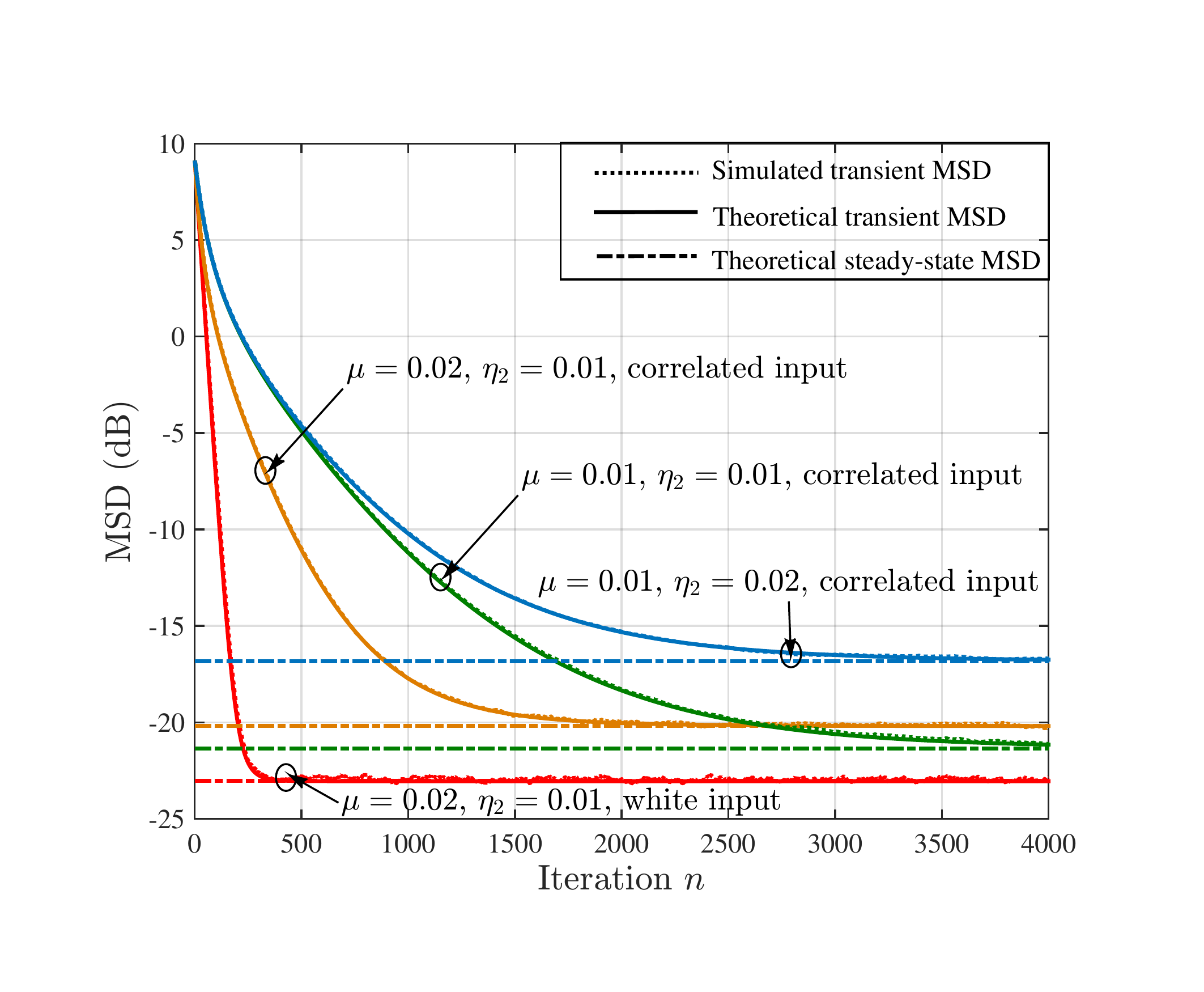}
   	\end{minipage}}  \hspace{2.5mm}
	\vspace{-2mm}
	\caption{Learning curves and model validation of  Algorithm 2 with different settings.}
	\label{fig:Alg2}
	\vspace{-8mm}
\end{figure*}

\begin{figure*}[!t]
	\hspace{-0cm}
	\subfigure[Using $\bTheta_1$.]{
 	\label{fig:Alg1THT1}
   	\begin{minipage}[c]{.3\linewidth}
   		\centering
      		\includegraphics[trim = 0mm 10mm 0mm 0mm, clip, scale=0.35]{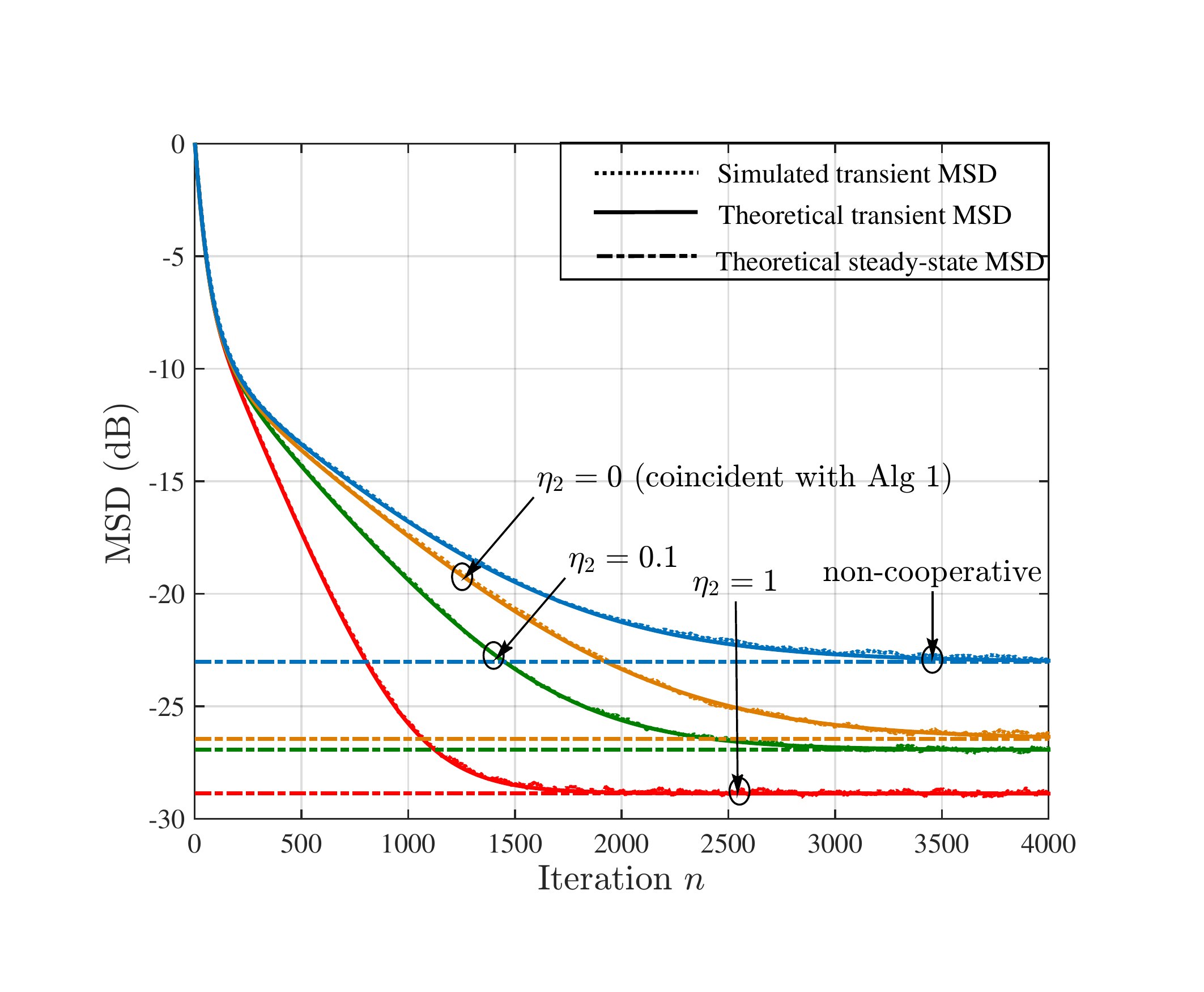}
   	\end{minipage}} \hspace{30mm}
	\subfigure[Using $\bTheta_2$.]{
 	\label{fig:Alg2THT1}
   		\begin{minipage}[c]{.3\linewidth}
      		\includegraphics[trim = 0mm 10mm 0mm 0mm, clip, scale=0.35]{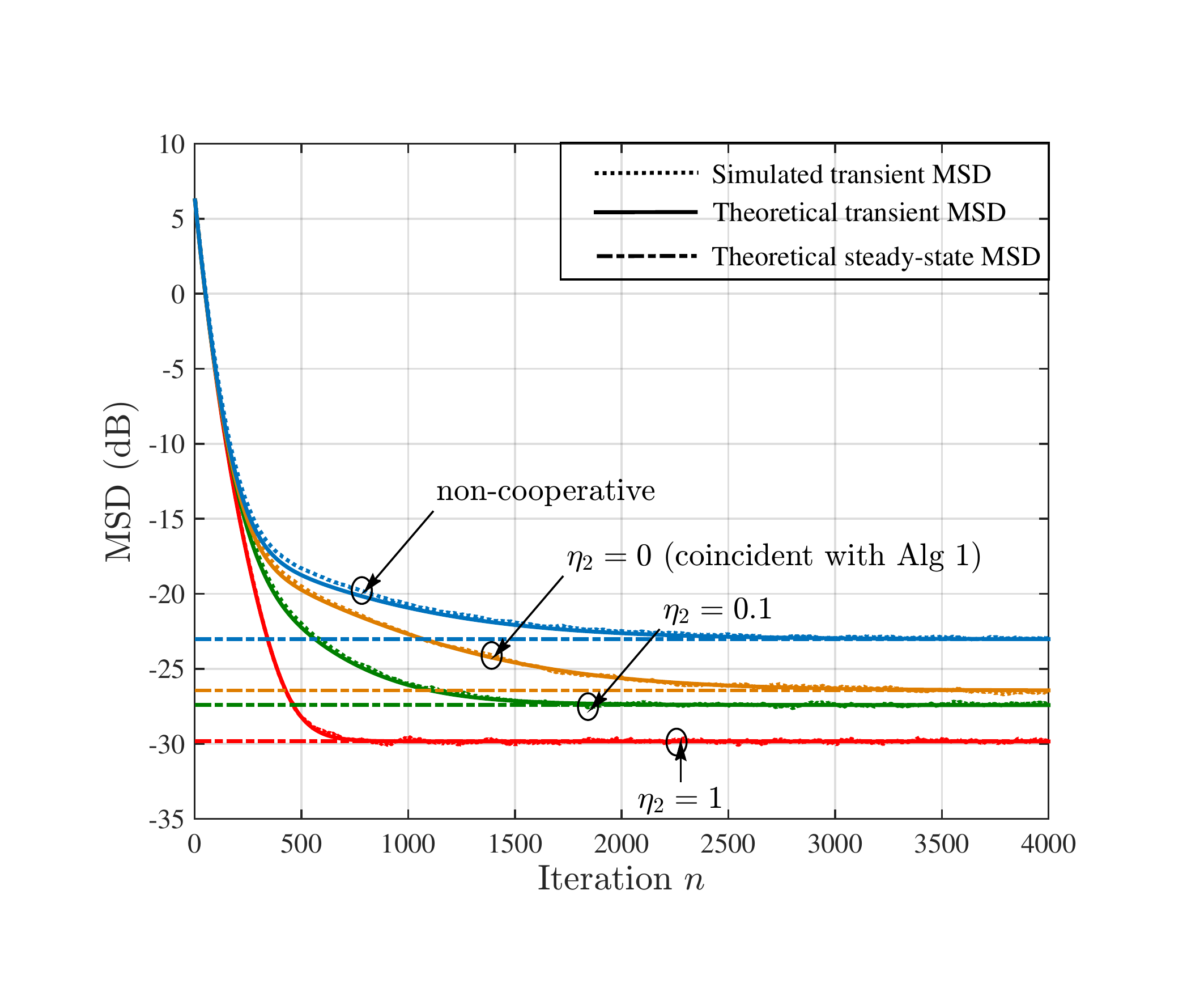}
   	\end{minipage}}  \hspace{2.5mm}
	\vspace{-2mm}
	\caption{Learning curves and model validation of  the algorithms using $\bxi_k$ with small variances.}
	\label{fig:AlgSV}
	\vspace{-8mm}
\end{figure*}

In the first setting, we assumed that model~\eqref{eq:structure} matches the observation data. The entries of the coefficient vectors $\bu^o$ and $\bxi_k^o$ were sampled from the Gaussian distribution $\cp{N}(0,1)$. The step-size parameter $\mu$ for Algorithm~1 was successively set to $0.01$ and $0.02$.  A uniform combination matrix $\cb{A}$ with $a_{\ell k} = |\N{k}|^{-1}$ was used. {With $\bTheta_1$, note that matrix $\bS_{\bTheta}$ is equal to $\bI_5$. With $\bTheta_2$, it was successively set to $\bTheta\bTheta^*+\bTheta_\perp^{\phantom{*}}\bTheta_\perp^*$ as in \eqref{eq:Alg1_2}, and to $\bI_5$.} The transient behavior and the steady-state MSD were determined theoretically. The results with subspace settings $\bTheta_1$ and $\bTheta_2$, for white and correlated input data, are shown in Fig.~\ref{fig:Alg1}. {It can be observed that setting $\bS_{\bTheta}$ to $\bI_5$ for $\bTheta_2$ leads to a better convergence behavior.} For Algorithm~2, we did not set the parameter $\eta_1$ explicitly but we used the same combination matrix~$\cb{A}$ as for Algorithm~1. Parameters $(\mu,\eta_2)$ were set to $(0.02,0.01)$ with white input data. With correlated input data, the following combinations $(\mu,\eta_2)$ were considered: $\{(0.01,0.01);(0.01,0.02);(0.02,0.01)\}$. The results are shown in Fig.~\ref{fig:Alg2}. The simulation results match the theoretical results, and illustrate the trade-off between the convergence speed and the steady-state MSD. It can also be observed with Algorithm 2 that a small value for $\eta_2$ is preferable since constraining the norm of node-specific components in the complementary subspace $\bTheta^\perp$ introduces a bias that can degrade the performance. As leaky-LMS, this kind of regularization can improve the stability of the algorithm for some particular problems and practical applications, at the cost of an extra estimation bias.  We then considered another scenario in order to illustrate the interest of the extra degree of freedom provided by $\eta_2$ in Algorithm 2. Experimental setups were left unchanged with correlated inputs except for the entries of $\bxi_k^o$, which were sampled from Gaussian distribution $\cp{N}(0, 0.01)$. We successively set $\eta_2$ to $0$, $0.1$ and $1$ in order to progressively constrain the variance of $\bxi_k$. Note that with $\eta_2=0$, Algorithm~2 reduces to Algorithm 1. The results with $\bTheta_1$ and $\bTheta_2$ are provided in Fig.~\ref{fig:AlgSV}. The result with non-cooperative LMS is also provided as a reference.

\begin{figure}[!t]
	\centering
 	\label{fig:Alg1THT2-ter}
   		\centering
      		\includegraphics[trim = 10mm 15mm 0mm 20mm, clip, scale=0.35]{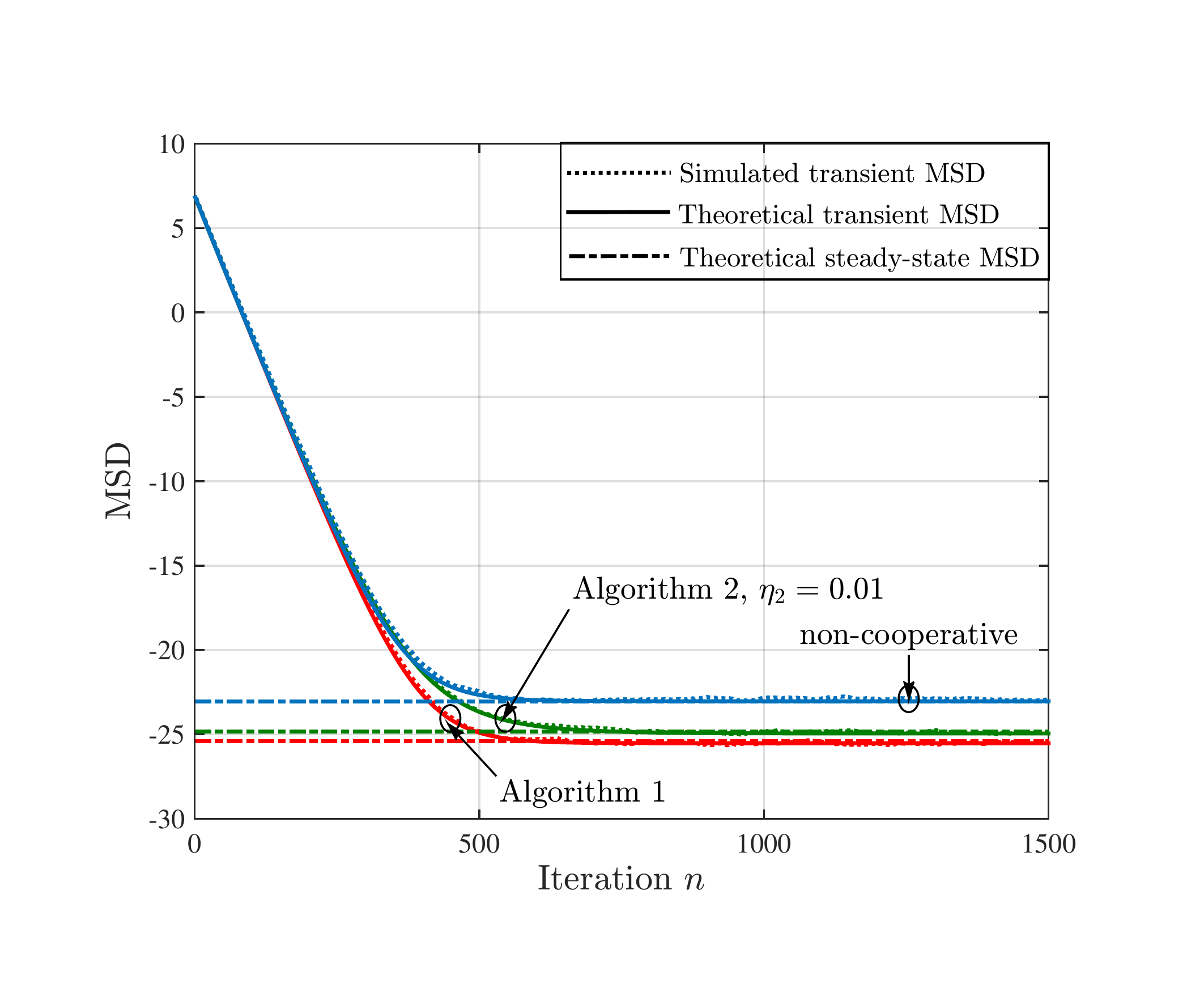}
	\vspace{-5mm}
	\caption{Learning curves of the algorithms using $\bxi_k$ with small variances.}
	\label{fig:ModelThtXi-ter}
	\vspace{-5mm}
\end{figure}

In the second setting, we assumed that the node-specific components $\beps^o_k$ in~\eqref{eq:structure} do not strictly lie in the complementary subspace $\bTheta^\perp$. To evaluate the robustness of our algorithms and {the power of the analytical models}, we set:
\begin{equation}
	\beps^o_k = \bTheta\cb{\nu}_k^o + \bThetac \bxi^o_k
\end{equation}
where $\cb{\nu}_k^o$ are zero-mean circular Gaussian variables. This setting refers to a non-ideal situation because components $\bTheta(\bu^o+\cb{\nu}_k^o)$ lie in $\text{span}(\bTheta)$ but differ from one node to another. The entries of $\bu^o$ and $\cb{\nu}_k^o$ were sampled from Gaussian distributions $\cp{N}(0,1)$ {and $\cp{N}(0,0.01)$}, respectively. The step-size $\mu$ was set to $0.01$ for Algorithms 1 and 2. Parameter $\eta_2$ in Algorithm 2 was set to $0.01$. {Subspace $\bTheta_1$ and white input signals were considered to test the model.} The transient behavior and the steady-state MSD were determined theoretically. The simulation results provided in Fig.~\ref{fig:ModelThtXi-ter} match the theoretical results, and illustrate that cooperation among nodes can still be beneficial when optimal solutions in the subspace $\bTheta$ are different but close to each other. This is another illustration of the conclusion reached in~\cite{Chen2015diffusion} for single-task diffusion LMS operating in multitask environments.

\begin{figure*}[!t]
	\hspace{-0cm}
	\centering
	\subfigure[Weight behavior of Algorithm~1.]{
 	\label{fig:Alg1THT1}
   	\begin{minipage}[c]{.3\linewidth}
   		\centering
      		\includegraphics[trim = 10mm 15mm 22mm 0mm, clip, scale=0.35]{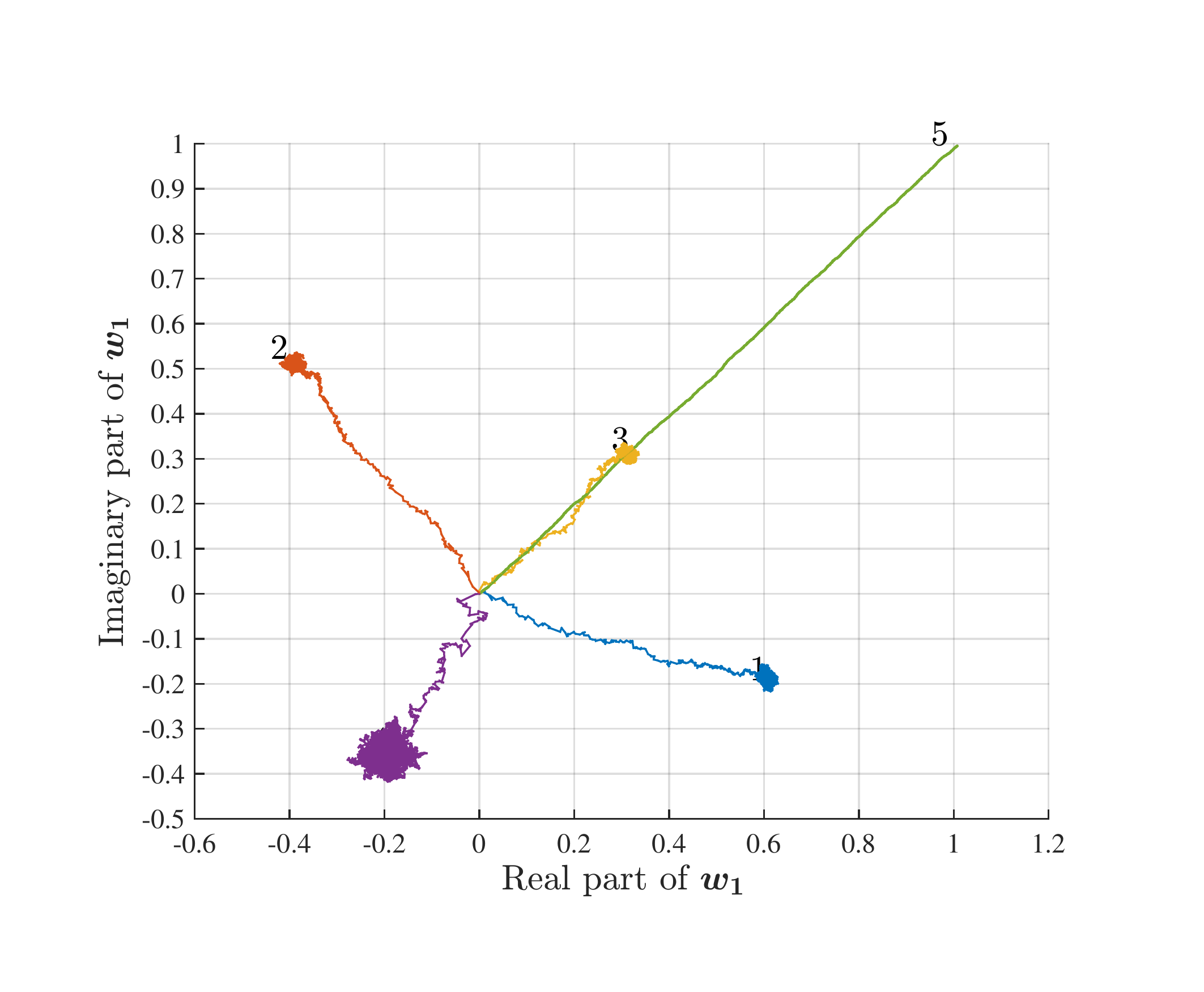}
   	\end{minipage}} \hspace{30mm}
	\subfigure[Weight behavior of Algorithm~2.]{
 	\label{fig:Alg2THT1}
   		\begin{minipage}[c]{.3\linewidth}
      		\includegraphics[trim = 10mm 15mm 22mm 0mm, clip, scale=0.35]{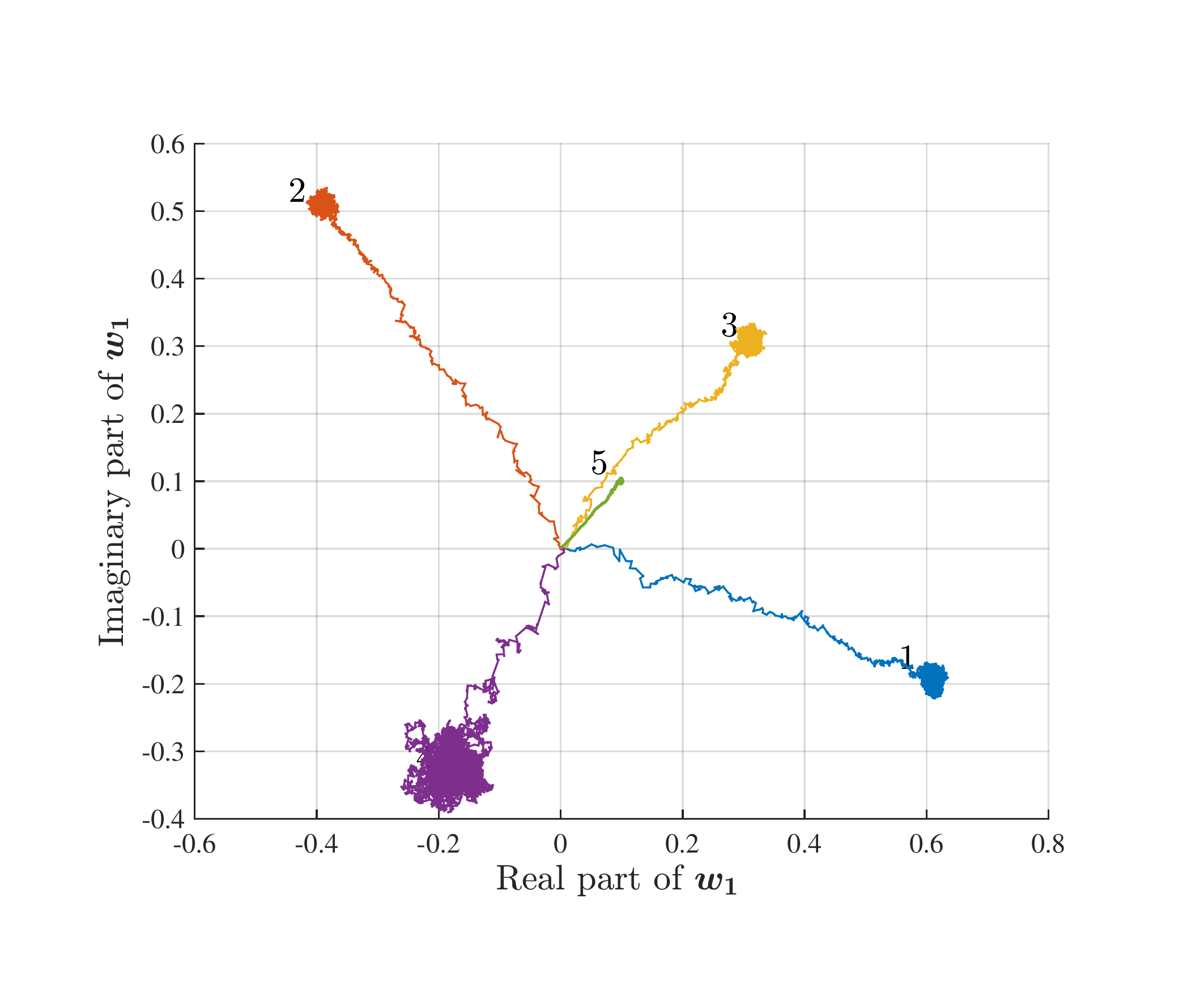}
   	\end{minipage}}  \hspace{2.5mm}
	\vspace{-2mm}
	\caption{Weight behavior of Algorithms~1 and~2 with singular inputs and non-zero mean disturbance.}
	\label{fig:drift}
\end{figure*}

{In the third setting, we exploited the leaky property of Algorithm~2 to promote its use in real applications.  It is well known that the (non-cooperative) leaky LMS algorithm introduces an estimation bias compared to the (non-cooperative)~LMS, but improves its robustness when applied to practical applications~\cite{sayed2003fundamentals}. In particular, it avoids the so-called weight-drift problem of the LMS algorithm~\cite{Nascimento1999unbiased}. To highlight this phenomenon in the context of diffusion adaptation, we assumed that, say, the last tap/channel of node $\#1$ was failing to work and was providing consistent null-valued readings, i.e., $[\bx_{n,1}]_5 = 0$ for all~$n$. We also assumed that, e.g., finite-precision effect was corrupting the combination step \eqref{eq:Alg1_2}, or~\eqref{eq:Alg2.b}, with an additive non-zero mean disturbance $\cb{q}_k$. The poor conditioning of regressors associated with a non-zero mean disturbance is known to possibly lead to a weight-drift problem. We considered the same experimental setup as in the first experiment with the standard basis $\bTheta_1$. We picked each entry of the random vectors~$\cb{q}_k$ according to the Gaussian distribution $\cp{N}(10^{-4},10^{-8})$. We set $\eta_2$ to $0.1$. All the vectors $\bw_k$ were initialized to $\cb{0}$. Fig.~\ref{fig:drift} shows the behavior of the weight vector at node $\#1$ for (a) Algorithm~1 with $\bS_{\Theta}=\bI_5$, and (b) Algorithm~2. We can observe the drift of the $5^\text{th}$ entry of $\bw_1$ with Algorithm~1. Algorithm~2 alleviates this effect. }

\subsection{Target localization}

We now consider a target localization problem. Cooperative localization with a diffusion strategy was already addressed in the case of a single target \cite{Sayed2013intr}, and of multiple nearby targets~\cite{Chen2014multitask}. We focus here on the case where targets lie in a manifold. 

\begin{figure}[!t]
\centering
	\subfigure[Network topology.]{
 	\label{fig:TopC}
   	\begin{minipage}[c]{.3\linewidth}
      		\includegraphics[trim = 30mm 30mm 10mm 40mm, clip, scale=0.3]{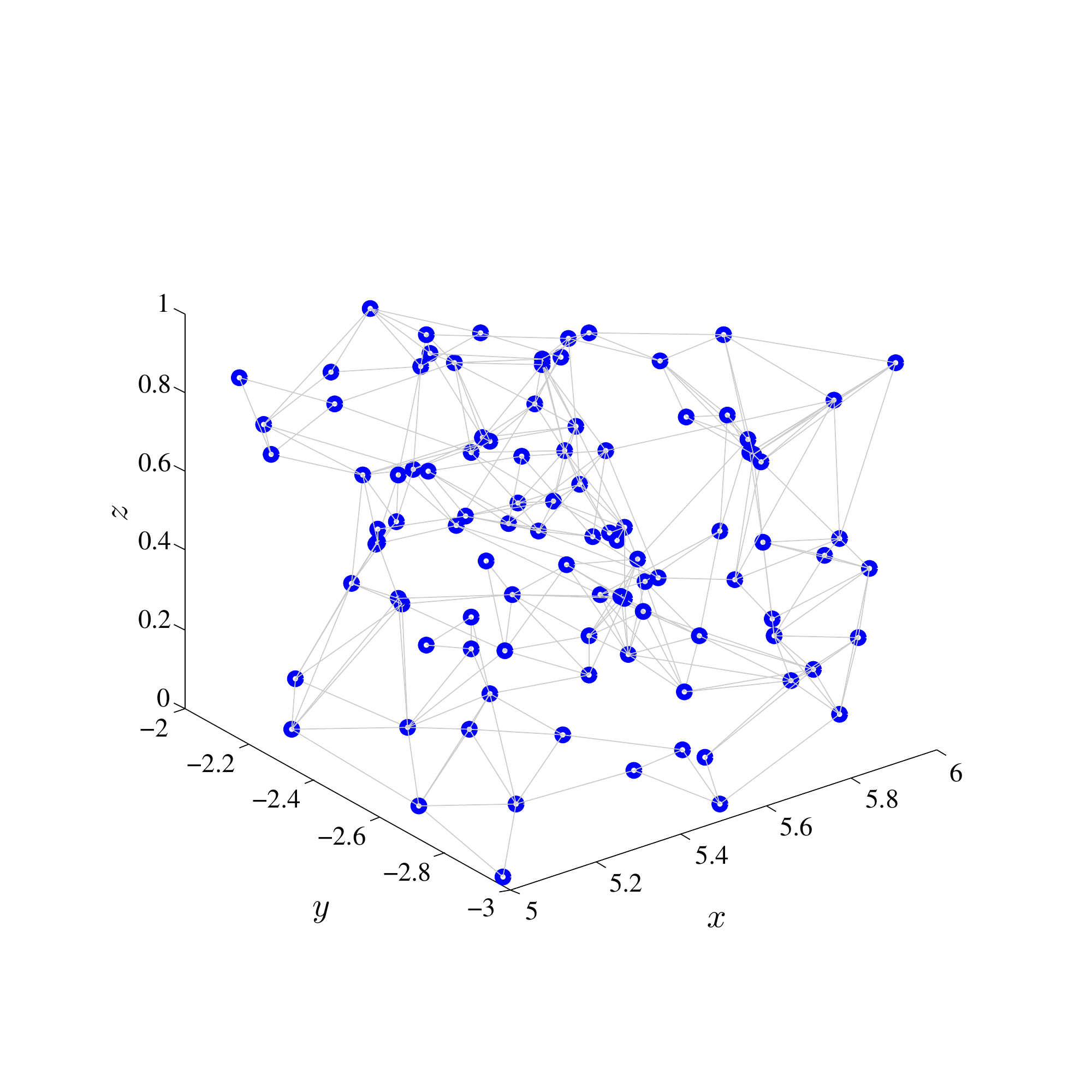}
   	\end{minipage}} 
	\hspace{2cm}
	\subfigure[Collinear targets.]{
 	\label{fig:Col}
   		\begin{minipage}[c]{.3\linewidth}
   		\centering
      		\includegraphics[trim = 0mm -25mm 0mm -10mm, clip, scale=0.4]{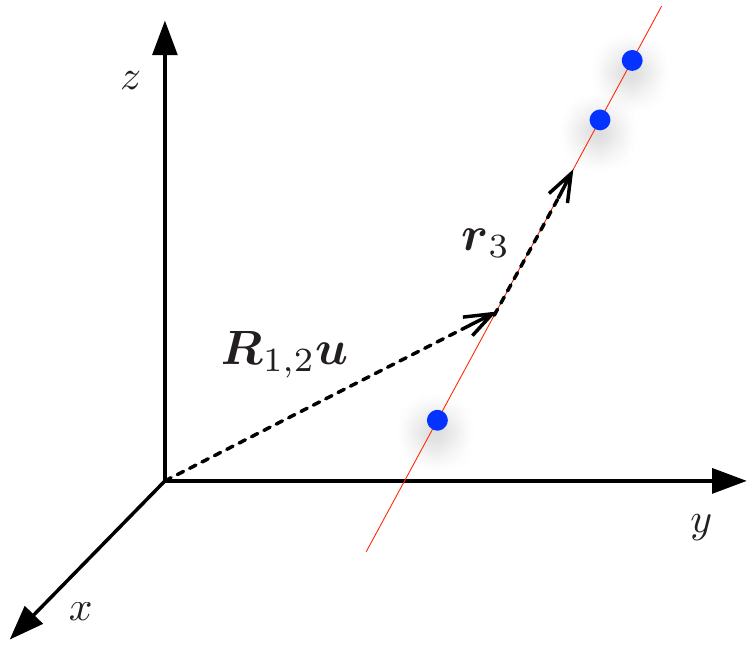}
   	\end{minipage}} \hfill
	\caption{Network topology and locations of targets.}
	\label{fig:Network_Col}
	\vspace{-5mm}
\end{figure}

\begin{figure*}[!t]
	\subfigure[Network MSD.]{
 	\label{fig:Col_MSD}
   	\begin{minipage}[c]{.3\linewidth}
   		\hspace{-2mm}
      		\includegraphics[trim = 20mm 10mm 0mm 10mm, clip, scale=0.35]{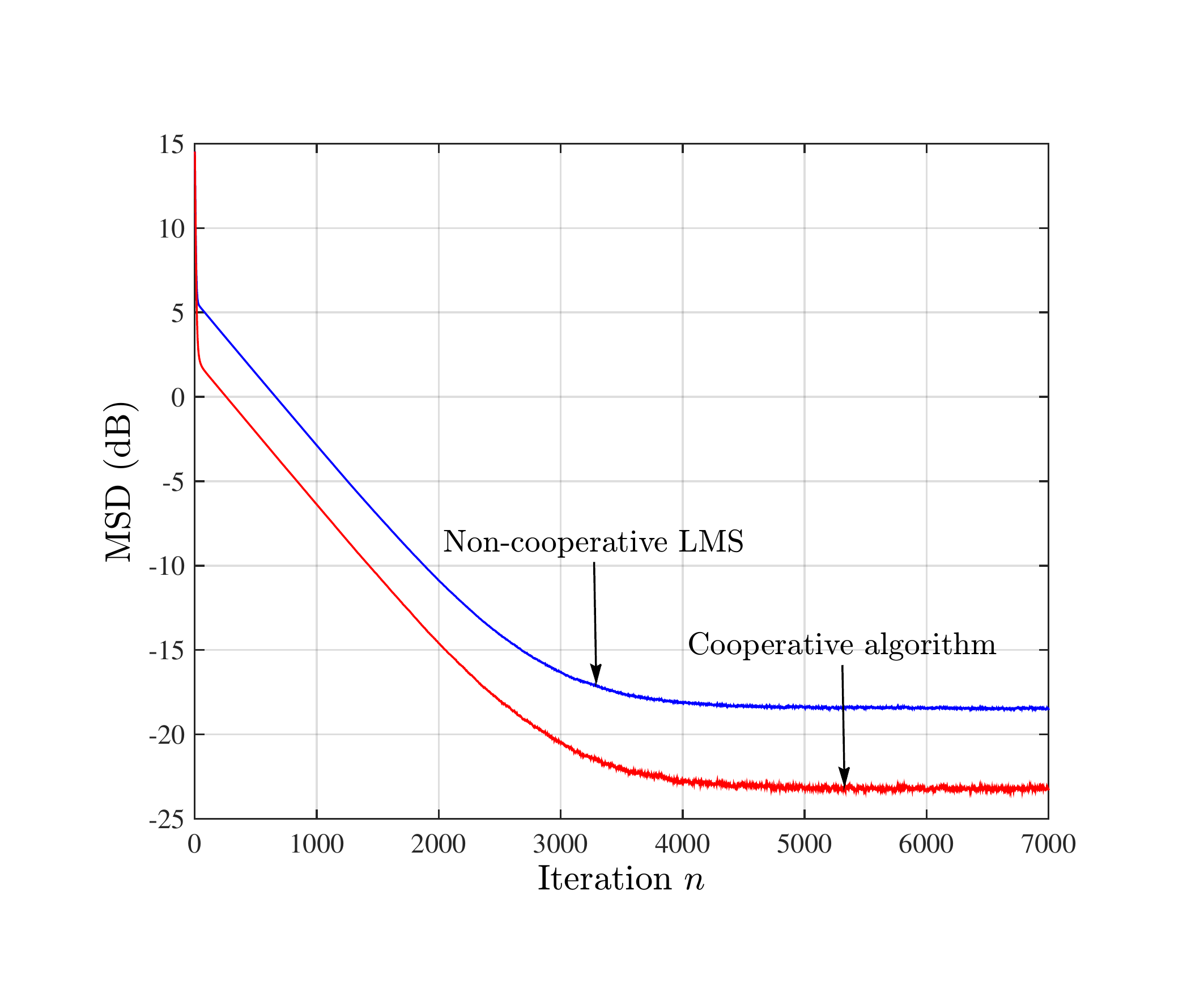}
   	\end{minipage}} \hfill
	\subfigure[Estimation results (non-coop. LMS).]{
 	\label{fig:Res_Non}
   		\begin{minipage}[c]{.3\linewidth}
		\hspace{-2mm}
      		\includegraphics[trim = 20mm 10mm 0mm 10mm, clip, scale=0.35]{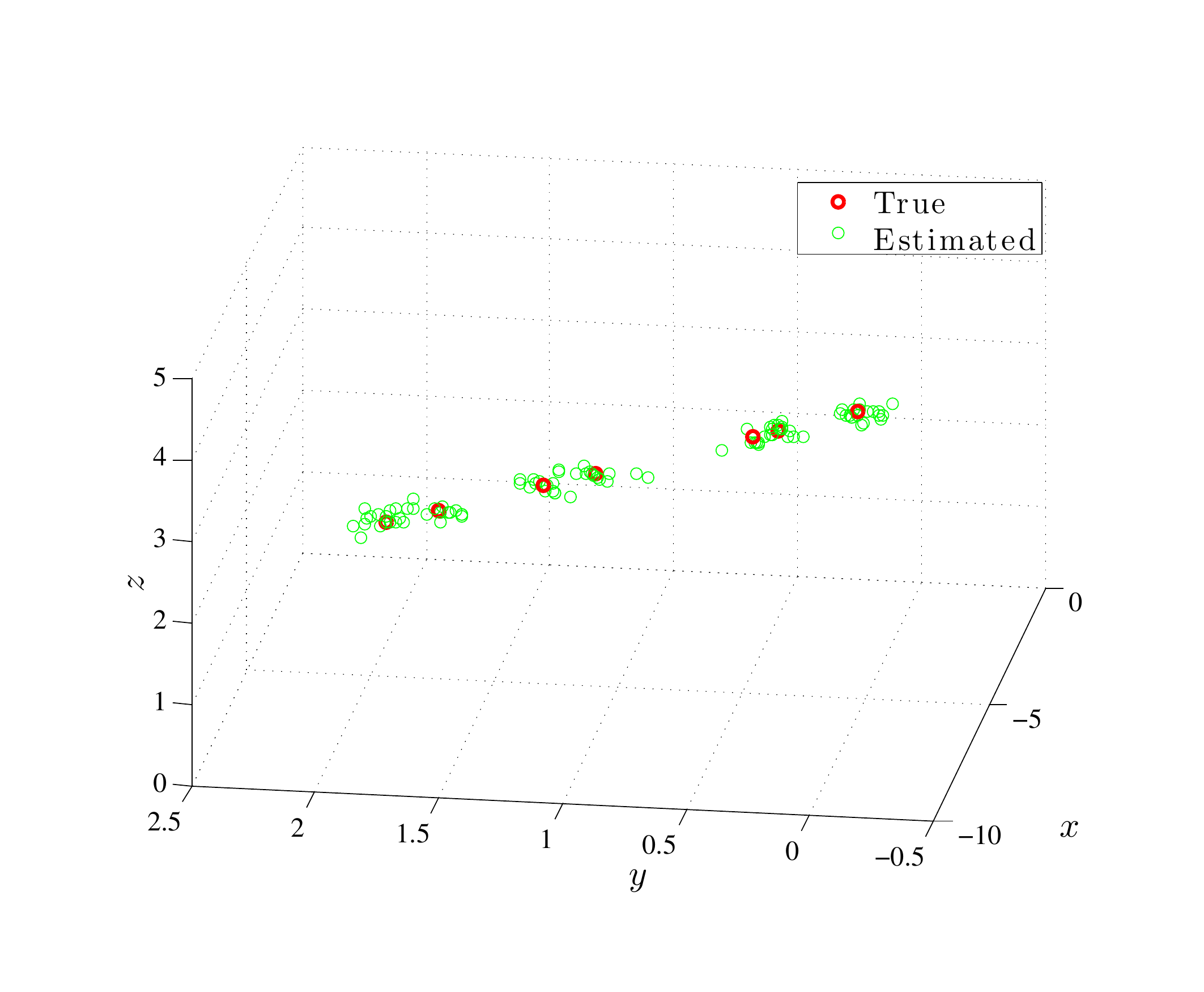}
   	\end{minipage}} \hfill
	\subfigure[Estimation results (Algorithm 1).]{
 	\label{fig:Res_Cop}
   		\begin{minipage}[c]{.3\linewidth}
   		\hspace{-2mm}
      		\includegraphics[trim = 20mm 10mm 0mm 10mm, clip, scale=0.35]{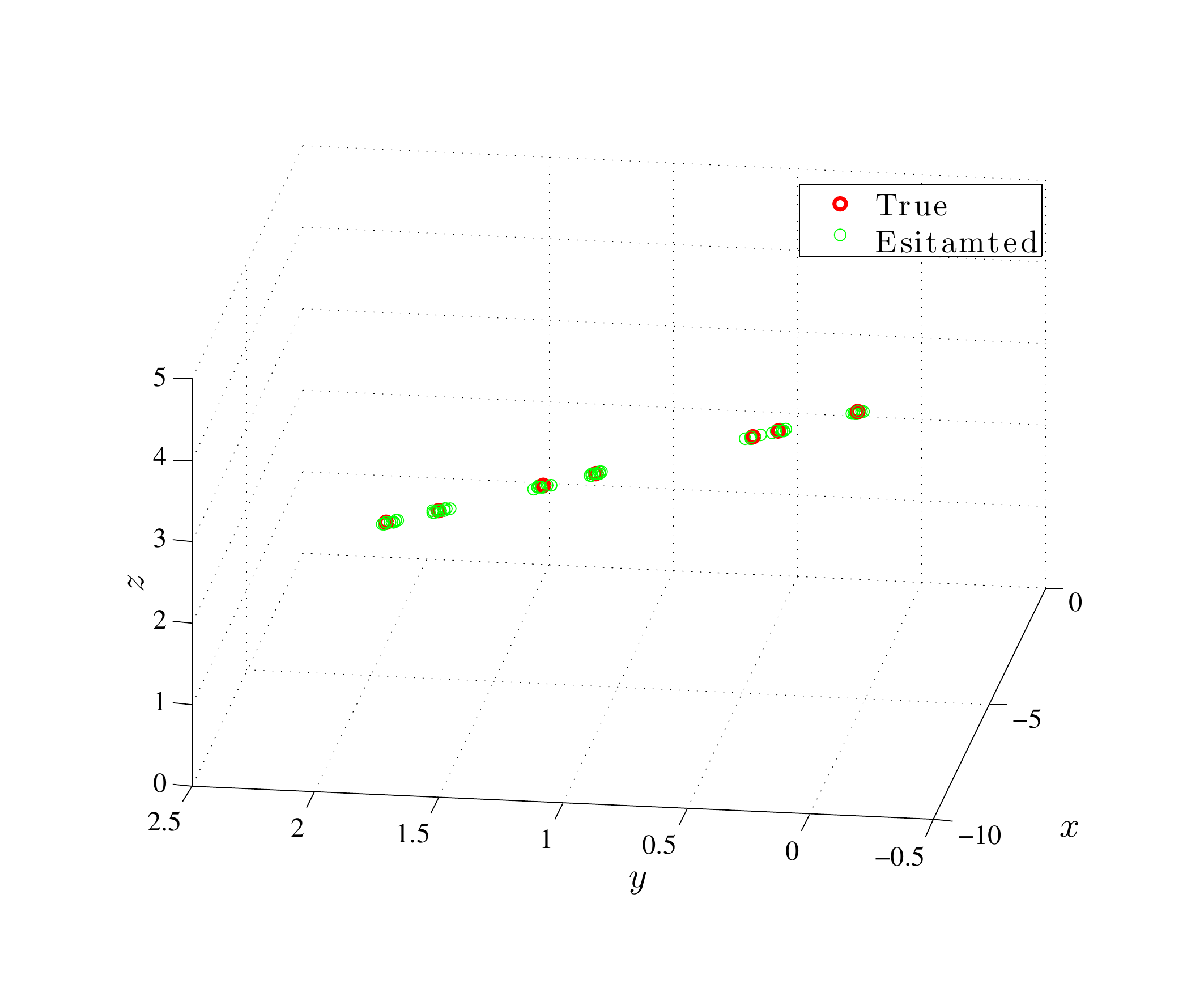}
   	\end{minipage}}
	\caption{Estimated network MSD and estimation results for a single realization.}
	\label{fig:Network_Col_MSD}
	\vspace{-5mm}
\end{figure*}

To make the presentation clearer, we assumed that the targets were collinear in $\R^3$. Their locations were estimated by the network with $100$ nodes shown in Fig.~\ref{fig:TopC}. Each node randomly selected a target to localize. Let $\cb{\cp{R}} $ be a member of the rotation group $\text{SO}(3)$ defined by the matrix $\bR = \bR_x(\theta_x)\,\bR_y(\theta_y)\,\bR_z(\theta_z)$, where $\bR_x(\theta_x)$, $\bR_y(\theta_y)$ and $\bR_z(\theta_z)$ are rotation matrices that rotate vectors by an angle of $\theta_{x,y,z}$ around $x$, $y$ and $z$ axis, respectively. 
The coordinate vector $\bw_{q}^o$ of each target~$q$ was generated as follows:
\begin{equation}
	\label{eq:locations}
	\bw_q^o = \bR_{1,2}\bu + \epsilon_q \br_{3}
\end{equation}
where $\bR_{1,2}$ is the matrix composed of the first and second columns of $\bR$, and $\br_{3}$ corresponds to the third column of~$\bR$. {As illustrated in Fig.~\ref{fig:Col}, this model means that all targets lie on a common line defined by point $\bR_{1,2}\bu$ and direction vector~$\br_3$. Parameter $\epsilon_q$ characterizes the location of each target $q$ on this line. We considered the problem of estimating~$\bu$ (common to all targets) and the parameters~$\epsilon_q$ for seven targets.} We set the angles and the parameter vectors in \eqref{eq:locations} as follows:
\begin{align}
	&\theta_x = \frac{\pi}{6}, \qquad \theta_y = \frac{\pi}{3}, \qquad \theta_z = \frac{\pi}{4} \\
	&\bv = [1~2]^\top \\
	&\epsilon_1 = 0, \epsilon_2 =1, \epsilon_3 =3, \epsilon_4 = 4, 
	\epsilon_5 =7, \epsilon_6 = 7.5, \epsilon_7 = 9
\end{align}
The distance between each agent $k$ and target $q$ can be expressed in the inner product form:
\begin{equation}
	r_{kq} = \bx_{kq}(\bw_q^o-\bp_k)
\end{equation}
where $\bp_k$ is the location of agent $k$, and $\bx_{kq}$ is the unit-norm row vector pointing from $\bp_k$ to $\bw_q^o$. We assumed that agents were aware of their location $\bp_k$. Let $d_{kq}=r_{kq}+\bx_{kq}\,\bp_k$, that is, $d_{kq} = \bx_{kq} \,\bw_q^o$. The problem was to estimate $\bw_q^o$ from noisy streaming measurements $\{d_{kq}(n),\bx_{kq,n}\}$ collected by each agent $k$, and governed by the linear model~\cite{Sayed2013intr}:
\begin{align}
	\label{eq:model-loc}
	&d_{kq}(n)	= \bx_{kq,n}\bw_q^o + z_{kq}(n) \nonumber\\
	&\text{with}\nonumber\\
	&\bx_{kq,n} = [1-\beta_{k}(n)]\,\bx_{kq} + \bx_{kq}^{\perp}\,\text{diag}\{\alpha_{k1}(n), \alpha_{k2}(n)\}
 \end{align}
%
%
with $z_{kq}(n)$ a zero-mean temporally and spatially i.i.d. Gaussian noise of variance $\sigma^2_z$. As shown in~\eqref{eq:model-loc}, the measured direction vector $\bx_{kq,n}$ was assumed to be a noisy realization of the unit-norm vector pointing from $\bp_k$ to $\bw^o_q$, with $\bx_{kq}^{\perp}$ a unit-norm orthogonal contribution to $\bx_{kq}$. Random variables $\alpha_{k1}(n)$, $\alpha_{k2}(n)$, $\beta_{k}(n)$  and $z_{k}(n)$ were zero-mean Gaussian with standard deviation $\sigma_{\alpha_1}=\sigma_{\alpha_2}=0.1$, $\sigma_\beta=0.001$ and $\sigma_z = 0.3$, respectively. We ran the (non-cooperative) LMS algorithm at each node, and Algorithm~1, with $\bTheta =\bR_{1,2}$ and $\bThetac = \br_{3}$. The step-size $\mu$ was set to $0.1$. A uniform  combination matrix $\cb{A}$ with $a_{\ell k} = |\N{k}|^{-1}$ was used for Algorithm 1, where $|\N{k}|$ denotes the cardinality of the neighborhood of node $k$. Figure~\ref{fig:Col_MSD} compares the MSD of these strategies. Figures~\ref{fig:Res_Non} and~\ref{fig:Res_Cop} show one realization of the target locations estimated with the (noncooperative) LMS algorithm and Algorithm 1. This experiment illustrates the advantage of cooperative strategies over the non-cooperative one.

\vspace{-2mm}
\section{Conclusion and perspectives}

In this paper, we formulated an online multitask adaptation problem that assumes that all tasks share a common latent feature representation, locally refined by node-specific contributions. This model can be extended into  interesting directions by imposing new constraints, depending on applications. Based on this principle, we derived two cooperative algorithms and analyzed their performance. Although this work considers that common representation subspaces are known a priori, it paves the way towards more general frameworks.

\begin{appendices}

\section{Proof of Lemma 1}
\label{app:proof-lemma1}

The uniqueness of the solution of~\eqref{eq:P1} follows from the strict convexity of~\eqref{eq:Jglob.perp}, which is  ensured by the positive definiteness of its Hessian matrix. {For the quadratic cost~\eqref{eq:Jglob.perp}, the Hessian matrix with respect to the vector of stacked variables $\col\{\bu, \bxi_1, \dots, \bxi_N\}$ is block diagonal~\cite[App. B]{Sayed2014adaptation}, with blocks given by the following matrix $\bX$ and its transpose:
\begin{align}
	\label{eq:hss}
	&\nabla^2 J^\text{glob} = 
	\left[\begin{array}{cc}
	\bX 	&  \cb{0}    \nonumber \\
	\cb{0} & \bX^\top
	\end{array}\right] \\
	& \noindent\text{with}\quad  \nonumber \\
	& \footnotesize \bX \!=\!
	\left(\begin{array}{c|ccc}
	\bTheta^* \big(\sum_{k=1}^N \bR_{x,k} \big) \bTheta 	&  \bTheta^*\bR_{x,1} \bThetac  
												& \dots   & \bTheta^*\bR_{x,N} \bThetac    \\
	\hline
	\bThetac^*\bR_{x,1} \bTheta						&  \bThetac^*\bR_{x,1} \bThetac  &   & {\large\cb{0}} \\
	\vdots                                                    				&	&  \ddots   &     \\
	\bThetac^*\bR_{x,N} \bTheta						&        {\large\cb{0}}	&      &    \bThetac^*\bR_{x,N} \bThetac
	\end{array}\right)
\end{align} 
where $\bTheta$ and $\bThetac$ have full column rank. The positive definiteness of \eqref{eq:hss} can be checked by verifying the positive definiteness of each term $\bTheta^*\bR_{x,k}\bTheta$ and of the Schur complement relative to the block diagonal corner of $\bX$, namely,~\cite{zhang2005schur}
\begin{equation}
         \label{eq:schur.hss}
         \begin{split}
       \text{Schur}(\bX) 
       = \sum_{k=1}^N \!\big[  \bTheta^*  \bR_{x,k} \bTheta \!-\! \bTheta^*\bR_{x,k} \bThetac (\bThetac^* \bR_{x,k} \bThetac)^{-1}\!\bThetac^*\bR_{x,k} \bTheta \big]
          \end{split}
\end{equation}
where each inverse $(\bThetac^* \bR_{x,k} \bThetac)^{-1}$ exists since $\bThetac$ has full column rank.}
{Each term inside} the summation \eqref{eq:schur.hss} is positive definite {since it is the Schur complement of the block $\bTheta^*  \bR_{x,k} \,\bTheta$} in the positive definite matrix:
\begin{equation}
                \left( \begin{array}{cc} \!\bTheta^*  \bR_{x,k} \bTheta  &  \bTheta^*\bR_{x,k} \bThetac\! \\  \!\bThetac^*\bR_{x,k} \bTheta & \bThetac^* \bR_{x,k} \bThetac\! \end{array}\right) \!=\! [\bTheta   \;\;   \bThetac]^*  \bR_{x,k}  [\bTheta \;\;  \bThetac] \!>\! 0
\end{equation}
{This guarantees the positive definiteness of~\eqref{eq:schur.hss}. It follows that the cost in~\eqref{eq:Jglob.perp} is strictly convex and has a unique minimizer.}

\section{Proof of Lemma 2}
\label{app:proof-lemma2}

Without loss of generality, assume that $\eta_1 > \eta_2$. Otherwise, replace \eqref{eq:Jglob.pert.regb} by: 
\begin{equation}
	\begin{split}
	J^\text{glob}&\big(\bu, \{\beps_k\}_{k=1}^N\big) = \sum_{k=1}^N {\E}\big\{|d_{k}(n) 
	- \bx_{k,n}(\bTheta\bu+\beps_k)|^2\big\}+ \eta_1\, \sum_{k=1}^N \|\beps_k\|^2 
	+ (\eta_2-\eta_1)\, \sum_{k=1}^N \|\bP_{\bThetac}\beps_k\|^2
	\end{split}
\end{equation}
{Recalling that $\bP_{\bThetac} = \bI_L - \bP_{\bTheta}$,} the objective function~\eqref{eq:Jglob.pert.reg} can be written as follows:
\begin{equation}
	\label{eq:Jglob.pert.regb}
	\begin{split}
	&J^\text{glob}\big(\bu, \{\beps_k\}_{k=1}^N\big) = (\eta_1-\eta_2) \sum_{k=1}^N \|\bP_{\bTheta}\beps_k\|^2+ \underbrace{\sum_{k=1}^N {\E}\big\{|d_{k}(n)  
	- {\bx_{k,n}(\bTheta\bu+\beps_k)}|^2\big\} + \eta_2 \sum_{k=1}^N \|\beps_k\|^2}_{J_1^\text{glob}}  
	\end{split}
\end{equation}
The uniqueness of the minimizer of \eqref{eq:Jglob.pert.reg} follows from its strict convexity. For the quadratic cost {in} \eqref{eq:Jglob.pert.regb}, the Hessian of $J_1^\text{glob}$ with respect to the vector of stacked variables $\col\{\bu, \beps_1, \dots, \beps_N\}$ {is again block diagonal, with its blocks determined by the matrix $\bY$ below and its transpose:
\begin{align}
	\label{eq:hss2}
	&\nabla^2 J_1^\text{glob} = 
	\left[\begin{array}{cc}
	\bY 	&  \cb{0} \\
	\cb{0} & \bY^\top
	\end{array}\right] \\
	&\text{with}\quad  \nonumber \\
	& \footnotesize \bY =
	\left(\begin{array}{c|ccc}
	\bTheta^* \big(\sum_{k=1}^N \bR_{x,k}\big) \bTheta  	& \bTheta^*\bR_{x,1}    & \dots   & \bTheta^*\bR_{x,N}     \\
	\hline
	\bR_{x,1} \bTheta	& \bR_{x,1}+\eta_2\,\bI 			&  		& {\large\cb{0}}	\\
	\vdots			&							& \ddots    & \\
	\bR_{x,N} \bTheta	& {\large\cb{0}}			&		& \bR_{x,N} +\eta_2\,\bI
	\end{array}\right)
\end{align}
The positive definiteness of \eqref{eq:hss2} can be checked by verifying the positive definiteness of each term $\bR_{x,k} +\eta_2\,\bI$ and {of the Schur complement relative to the right block diagonal corner} in \eqref{eq:hss2}, namely,~\cite{zhang2005schur}}
\begin{align}
	\label{eq:schur.hss2}
	\text{Schur}(\bY)  
       	= \sum_{k=1}^N \! \big[\bTheta^*  \bR_{x,k}\bTheta \!-\! \bTheta^*\bR_{x,k} (\bR_{x,k} 
	+\eta_2\bI)^{-1}\bR_{x,k} \bTheta\big]
\end{align}
Since they are positive definite, {each covariance matrix} $\bR_{x,k}$ can be decomposed as follows:
\begin{equation}
	\bR_{x,k} = \bU_k\, \text{diag}\{\lambda_{k,1}, \dots, \lambda_{k,L}\} \, \bU_k^*
\end{equation}
where the $\lambda_{k,i}$ are the eigenvalues of $\bR_{x,k}$, which are real and positive, and $\bU_k$ is the corresponding matrix of eigenvectors. Since $\bU_k$ is an orthonormal matrix, each term in the summation~\eqref{eq:schur.hss2} can be written as:
\begin{equation}
	\begin{split}
	&\bTheta^*  \bR_{x,k}\bTheta - \bTheta^*\bR_{x,k} (\bR_{x,k} + \eta_2\bI)^{-1}\bR_{x,k} \bTheta \\
	=& \bTheta^*\bU \text{diag}\left\{\lambda_{k,1}
	- \mathsmaller{\frac{\lambda_{k,1}^2}{\lambda_{k,1}+\eta_2}},\dots, \lambda_{k,L}
	- \mathsmaller{ \frac{\lambda_{k,L}^2}{\lambda_{k,L}+\eta_2}} \right\} \bU^* \bTheta > 0
           \end{split}
\end{equation}
Since $\bTheta$ has full column rank, the above matrix and the Schur complement~\eqref{eq:schur.hss2} are positive definite. In addition, the block diagonal matrix $\text{diag}\{\bR_{x,1}+\eta_2\bI, \dots, \bR_{x,N}+\eta_2\bI\}$ is positive definite. Finally, since $(\eta_1-\eta_2)\, \sum_{k=1}^N \|\bP_{\bTheta}\beps_k\|^2$ in~\eqref{eq:Jglob.pert.regb} is convex, problem~\eqref{eq:Jglob.pert.regb} is strictly convex and problem~\eqref{eq:P2} has a unique solution.

\end{appendices}

\bibliographystyle{IEEEbib}
\bibliography{reference}

\end{document}